\documentclass[oneside,australian,english]{siamart0516}
\usepackage[T1]{fontenc}
\usepackage[latin9]{inputenc}
\usepackage{array}
\usepackage{refstyle}
\usepackage{mathrsfs}
\usepackage{amssymb}
\usepackage{graphicx}

\makeatletter

\AtBeginDocument{\providecommand\secref[1]{\ref{sec:#1}}}
\AtBeginDocument{\providecommand\figref[1]{\ref{fig:#1}}}
\AtBeginDocument{\providecommand\subsecref[1]{\ref{subsec:#1}}}
\providecommand{\tabularnewline}{\\}
\RS@ifundefined{subsecref}
  {\newref{subsec}{name = \RSsectxt}}
  {}
\RS@ifundefined{thmref}
  {\def\RSthmtxt{theorem~}\newref{thm}{name = \RSthmtxt}}
  {}
\RS@ifundefined{lemref}
  {\def\RSlemtxt{lemma~}\newref{lem}{name = \RSlemtxt}}
  {}

\newcommand\eqref[1]{(\ref{#1})}

\usepackage{algpseudocode}
\usepackage{pgfplots}
\usepackage{mathtools}

\renewcommand{\figref}{\Figref}
\pgfplotsset{compat=1.3}
\usetikzlibrary{patterns}
\usetikzlibrary{plotmarks}

\@ifundefined{showcaptionsetup}{}{%
 \PassOptionsToPackage{caption=false}{subfig}}
\usepackage{subfig}
\makeatother

\usepackage{babel}
\usepackage{listings}
\addto\captionsaustralian{}
\addto\captionsenglish{}

\begin{document}

\title{High-Performance Tensor Contraction without Transposition}

\author{Devin A. Matthews\thanks{Institute for Computational Engineering and Sciences, the University of Texas at Austin, Austin, TX 78712, USA (\email{dmatthews@utexas.edu})}}
\maketitle
\begin{abstract}
Tensor computations\textendash in particular tensor contraction (TC)\textendash are
important kernels in many scientific computing applications. Due to
the fundamental similarity of TC to matrix multiplication (MM) and
to the availability of optimized implementations such as the BLAS,
tensor operations have traditionally been implemented in terms of
BLAS operations, incurring both a performance and a storage overhead.
Instead, we implement TC using the flexible BLIS framework, which
allows for transposition (reshaping) of the tensor to be fused with
internal partitioning and packing operations, requiring no explicit
transposition operations or additional workspace. This implementation,
TBLIS, achieves performance approaching that of MM, and in some cases
considerably higher than that of traditional TC. Our implementation
supports multithreading using an approach identical to that used for
MM in BLIS, with similar performance characteristics. The complexity
of managing tensor-to-matrix transformations is also handled automatically
in our approach, greatly simplifying its use in scientific applications.
\end{abstract}


\section{Introduction}

Tensors are an integral part of many scientific disciplines \cite{vasilescu_multilinear_2002,smilde_multi-way_2005,bartlett_coupled-cluster_2007,kroonenberg_applied_2008,kolda_tensor_2009}.
At their most basic, tensors are simply a multidimensional collection
of data (or a multidimensional array, as expressed in many programming
languages). In other cases, tensors represent multidimensional transformations,
extending the theory of vectors and matrices. The logic of handling,
transforming, and operating on tensors is a common task in many scientific
codes, often being reimplemented many times as needed for different
projects or even within a single project. Calculations on tensors
also often account for a significant fraction of the running time
of such tensor-based codes, and so their efficiency has a significant
impact on the rate at which the resulting scientific advances can
be achieved.

In order to perform a tensor contraction, there are currently two
commonly used alternatives: (1) write explicit loops over the various
tensor indices (this is the equivalent of the infamous triple loop
for matrix multiplication), or (2) ``borrow'' efficient routines
from optimized matrix libraries such as those implementing the BLAS
interface \cite{blas1,blas2,blas3}. Choice (1) results in a poorly-performing
implementation if done naively due to the lack of optimizations for
cache reuse, vectorization, etc. as are well-known in matrix multiplication
(although for a high-performance take on this approach see GETT \cite{gett}
as discussed in \secref{Related-Work}), and also generally requires
hardcoding the specific contraction operation, including the number
of indices on each tensor, which indices are contracted, in which
order indices are looped over, etc. This means that code cannot be
efficiently reused as there are \emph{many }possible combinations
of number and configuration of tensor indices. Choice (2) may provide
for higher efficiency, but requires mapping of the tensor operands
to matrices which involves both movement of the tensor data in memory
and often this burden falls on the user application. Alternatively,
the tensors may be \emph{sliced} (i.e. low-dimensional sub-tensors
are extracted) and used in matrix multiplication, but this again has
drawbacks as discussed below.

Several libraries exist that take care of part of this issue, namely
the tedium of encoding the logic for treating tensors as matrices
during contractions. For example, the MATLAB Tensor Toolbox \cite{ttoolbox}
provides a fairly simple means for performing tensor contractions
without exposing the internal conversion to matrices. The NumPy library
\cite{NumPy} provides a similar mechanism in the Python language.
Both libraries still rely on the mapping of tensors to matrices and
the unavoidable overhead in time and space thus incurred. In many
cases, especially in languages common in high-performance computing
such as FORTRAN and C, a pre-packaged solution is not available and
the burden of implementing tensor contraction is left to the author
of the scientific computing application, although several libraries
for tensor contraction in C++ have recently been developed \cite{libtensor,ctf,tiledarray1,tiledarray2}.

A natural solution to this problem is to create a dedicated tensor
library implementing a high-performance, ``native'' tensor contraction
implementation (without the use of the BLAS), but with similar optimizations
for cache reuse, vectorization, etc. as those used for matrix multiplication.
For a particular instance of a tensor contraction this is a feasible
if not tedious task, using existing techniques from dense linear algebra.
However, making a \emph{general} run-time tensor library is another
proposition altogether, as it is highly desirable for such a library
to handle any number of tensor indices, as well as number, order,
and position of contracted indices. Static code transformation/generation
techniques such as in Built-to-Order BLAS \cite{bto}, DxTer \cite{dxter},
and GETT \cite{gett} can produce highly efficient code but are much
less flexible since they require tensor dimensionality and ordering
(and in many cases, sizes) to be fixed at compile time. Since the
number of possible contraction types grows exponentially with the
number of indices, explicitly treating each instance of tensor contraction
is an uphill endeavor in large applications with many tensor contraction
instances. Not requiring a code-generation stage is also highly beneficial
to rapid prototyping of new algorithms that use tensor contraction,
while having a high-performance general tensor contraction library
can often turn prototypes into useful production code.

The newly-developed BLIS framework \cite{blis1,blis-multi,blis2}
implements matrix operations with a very high degree of efficiency,
including a legacy BLAS interface. However, unlike common BLAS libraries,
especially commercial libraries such as Intel's Math Kernel Library,
BLIS exposes the entire internal structure of algorithms such as matrix
multiplication, down to the level of the \emph{micro-kernel}, which
is the only part that must be implemented in highly-optimized assembly
language. In addition to greatly reducing the effort required to build
a BLAS-like library, the structure of BLIS allows one to view a matrix
operation as a collection of independent pieces. Matrix multiplication
(and other level 3 BLAS operations) are the default algorithms provided
by BLIS, but one may also use the available pieces to build a custom
framework for purposes beyond traditional dense linear algebra, examples
of which are given in \cite{chenhan} and \cite{FLAWN79}. We will
illustrate exactly how the flexibility of this approach may be used
to implement high-performance tensor contraction by breaking through
the traditionally opaque matrix multiplication interface. The benefits
of this approach are detailed in \secref{Results}, where our new
algorithm, Block-Scatter-Matrix Tensor Contraction (BSMTC), is compared
to a traditional tensor contraction approach, Transpose-Transpose-{\tt GEMM}-Transpose
(TTGT) as implemented by the Tensor Toolbox.

The specific contributions of this paper are:
\begin{itemize}
\item A novel logical mapping from general tensor layouts to a non-standard
(neither row-major nor column-major) matrix layout.
\item Implementations of key BLIS kernels using this novel matrix layout,
eliminating the need for explicit transposition of tensors while retaining
a matrix-oriented algorithm.
\item A new BLIS-like framework incorporating these kernels that achieves
high performance for tensor contraction and does not require external
workspace.
\item Efficient multithreading of tensor contraction within the aforementioned
framework.
\end{itemize}

\section{The Big Idea: Tensors As Matrices}

Before developing the theory of matrix multiplication and tensor contraction
in detail, let us first examine the high-level concepts which guide
our implementation of tensor contraction. First, we may introduce
tensors quite simply as the multi-dimensional extension of scalars
(0-dimensional tensors), vectors (1-dimensional tensors), and matrices
(2-dimensional tensors), or similarly as multi-dimensional arrays.
Tensor contraction is the also natural extension of matrix multiplication
to tensors, where for each index (dimension) in the two input and
the single output matrix we substitute one or more tensor dimensions.
Thus, a contraction of a 7-dimensional tensor $\mathscr{A}$ by a
5-dimensional tensor $\mathscr{B}$ into a 6-dimensional tensor $\mathscr{C}$
is in fact mathematically a matrix multiplication in disguise if we
say that for example, $\mathscr{A}$ is in fact 4-dimensional by 3-dimensional
(for some partition of the 7 dimensions into groups of 4 and 3), $\mathscr{B}$
is 3-dimensional by 2-dimensional, and $\mathscr{C}$ is merely 4-dimensional
by 2-dimensional. The problem with mapping tensor contraction to matrix
multiplication in practice is that the data layout of tensors in memory
is much more flexible and varied than for matrices, and a direct mapping
is generally not possible.

Since matrix multiplication and tensor contraction are really the
same thing, then, the obvious question is, ``how can existing high-performance
matrix multiplication algorithms be used for tensor contraction?''
The trick is to relate the single matrix dimensions, which are usually
described by a length (number of elements), and a stride (distance
between elements), to a whole group of tensor dimensions (especially
when the number of tensor dimensions is not known beforehand). As
we will show in the following sections, this relationship between
matrices and tensors is cleanly and efficiently accomplished by creating
a specialized matrix layout instead of simple scalar strides, which
by design describes the physical data layout of a specified tensor.
An existing matrix multiplication algorithm can then use this layout
to access matrix elements without knowing anything about tensors or
tensor contraction. We will demonstrate that this approach yields
a relatively simple and highly efficient tensor contraction algorithm.

\section{Matrix Multiplication\label{sec:Matrix-Multiplication}}

\begin{figure}
\begin{centering}
\includegraphics[width=0.6\columnwidth]{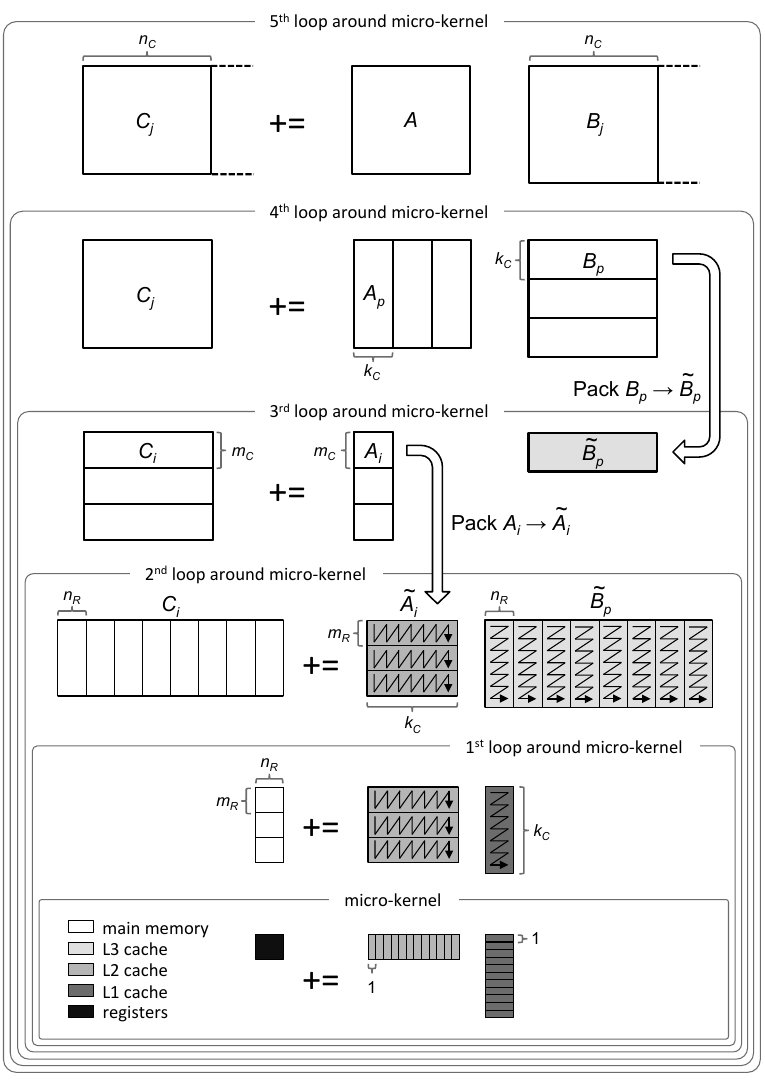}
\par\end{centering}
\caption{\label{fig:blis}The structure of a matrix multiplication operation
using the BLIS approach. Figure from https://github.com/flame/blis/wiki/Multithreading,
used with permission.}

\end{figure}

Let us first review the basic techniques used in high-performance
matrix multiplication, expressed generally as $C\coloneqq\alpha AB+\beta C$.
The matrices $A$, $B$, and $C$ are $m\times k$, $k\times n$,
and $m\times n$ respectively. We refer to the length and width of
each matrix as its \emph{shape}, and refer to individual elements
of each matrix as $A_{ip}$, $C_{ij}$, etc. where $0\le i<m$, $0\le j<n$,
and $0\le p<k$.\emph{ }Written element-wise, the basic matrix multiplication
operation is then $C_{ij}\coloneqq\alpha\sum_{p=0}^{k-1}A_{ip}\cdot B_{pj}+\beta C_{ij}\,\forall\,(i,j)\in m\times n$.
In this context, $m\times n$ is a shorthand for the Cartesian product
of the ranges $[0,m)$ and $[0,n)$, i.e. $m\times n=\{(i,j)\,|\,0\le i<m\,\vee\,0\le j<n\}$.
The use of Cartesian products over integers to denote such sets will
be used extensively. This element-wise definition of matrix multiplication
is critical to relating it to tensor contraction, although for brevity
we will assume $\alpha=1$ and $\beta=0$ henceforth.

Most current high-performance implementations follow the approach
pioneered by Goto \cite{goto1,goto2}, in which the matrices are successively
partitioned (sub-divided) into panels (sub-matrices), labeled for
example $A_{i}$, $B_{p}$, and $C_{j}$ in \figref{blis}, which
are designed to fit into the various levels of the cache hierarchy.
The panels of $A$ and $B$ are packed (copied) into temporary buffers
$\tilde{A}_{i}$ and $\tilde{B}_{p}$ in a special storage format
to facilitate vectorization and memory locality. The size of these
panels are determined by \emph{cache blocking parameters }$m_{C}$,
$n_{C}$, and $k_{C}$, such that $\tilde{A}_{i}$ (which stores an
$m_{C}\times k_{C}$ panel of $A$) is retained in the L2 cache while
$\tilde{B}_{p}$ (which stores a $k_{C}\times n_{C}$ panel of $B$)
is retained in the L3 cache (if present). In the original Goto approach,
these ``pack buffers'' are then fed into an \emph{inner kernel}
which performs the actual matrix multiplication sub-problem, and which
is typically written in hand-optimized assembly code.

The BLIS approach \cite{blis1} implements the \emph{inner kernel
}instead in terms of a much smaller and easier to write \emph{micro-kernel.
}In this approach, the panels of $A$ and $B$ as stored in $\tilde{A}_{i}$
and $\tilde{B}_{p}$ are further partitioned according to \emph{register
block sizes }$m_{R}$ and $n_{R}$ such that a pair of $m_{R}\times k_{C}$
and $k_{C}\times n_{R}$ \emph{slivers} of $\tilde{A_{i}}$ and $\tilde{B}_{p}$,
respectively, fit into the L1 cache. The micro-kernel then uses these
slivers to update a small $m_{R}\times n_{R}$ \emph{micro-tile }of
$C$, which is maintained in machine registers. These five cache and
register block sizes give rise to five loops (written in C or another
relatively high-level language) around the micro-kernel, as illustrated
in \figref{blis}.

The major logical and computational operations involved in the BLIS
approach are then: (1) partitioning of matrix operands, (2) packing
of matrix panels (as a set much smaller matrix slivers) into specially-formatted
buffers, and (3) invocation of the micro-kernel. We will see these
operations can be implemented in the context of general tensors instead
of matrices in later sections, in order to make use of the highly-tuned
BLIS framework for tensor contraction.

\section{Tensor Contraction}

A general $d$-dimensional tensor $\mathscr{A}\in\mathbb{R}^{n_{u_{0}}\times\ldots\times n_{u_{d-1}}}$
is defined as the set of scalar elements indexed by the set of indices
$u_{0}\ldots u_{d-1}$,

\begin{equation}
\mathscr{A}\equiv\{\mathscr{A}_{u_{0}\ldots u_{d-1}}\in\mathbb{R}\,|\,(u_{0},\ldots,u_{d-1})\in n_{u_{0}}\times\ldots\times n_{u_{d-1}}\}
\end{equation}
Often, the indices of a tensor ($u_{0},\ldots,u_{d-1}$) will be given
more convenient labels, such as in $\mathscr{A}_{abc\ldots}$, while
the length of the corresponding tensor dimensions will be denoted
$n_{a}$, $n_{b}$, etc. Indices which have the same label in more
than one tensor in any tensor expression must share the same value,
and the lengths of the corresponding dimensions in each tensor must
be identical. For example, in an expression such as $\mathscr{A}_{cfbd}\cdot\mathscr{B}_{fea}$,
the length of the index $f$, $n_{f}$, must be the same in $\mathscr{A}$
and $\mathscr{B}$, and only pairs of elements with the same value
of $f$ in $\mathscr{A}$ and $\mathscr{B}$ enter the expression.

Tensor contraction generalizes the concept of matrix multiplication
to higher dimensions (numbers of indices), just as tensors generalize
the notion of matrices. Given two input tensors $\mathscr{A}$ and
$\mathscr{B}$ of dimension $d(\mathscr{A})$ and $d(\mathscr{B})$,
respectively, and an output tensor $\mathscr{C}$ of dimension $d(\mathscr{C}$),
a tensor contraction is specified by first selecting ordered sets
of $d_{P}=(d(\mathscr{A})+d(\mathscr{B})-d(\mathscr{C}))/2$ dimensions
each from $\mathscr{A}$ and from $\mathscr{B}$ and labeling the
indices of the dimensions in both sets as $p_{0}\ldots p_{d_{P}-1}$.
Since the indices are labeled the same in both tensors, their values
must be the same (i.e. the indices are \emph{bound}). Similarly, the
remaining dimensions of $\mathscr{A}$ and an ordered set of $d_{I}=(d(\mathscr{A})+d(\mathscr{C})-d(\mathscr{B}))/2$
dimensions in $\mathscr{C}$ have their indices labeled $i_{0}\ldots i_{d_{I}-1}$.
Finally, the remaining $d_{J}=(d(\mathscr{B})+d(\mathscr{C})-d(\mathscr{A}))/2$
dimensions in $\mathscr{B}$ and $\mathscr{C}$ are arranged in a
selected relative order and their indices labeled $j_{0}\ldots j_{d_{J}-1}$.
The tensor contraction operation is then given element-wise by,

\begin{eqnarray}
\mathscr{C}_{\pi_{C}(i_{0}\ldots i_{d_{I}-1}j_{0}\ldots j_{d_{J}-1})}\coloneqq\underset{p_{0},\ldots,p_{d_{P}-1}=0}{\sum^{n_{p_{0}}-1,\ldots,n_{pd_{P}-1}-1}}\mathscr{A}_{\pi_{A}(i_{0}\ldots i_{d_{I}-1}p_{0}\ldots p_{d_{P}-1})}\cdot\mathscr{B}_{\pi_{B}(p_{0}\ldots p_{d_{P}-1}j_{0}\ldots j_{d_{J}-1})}\nonumber \\
\forall(i_{0},\ldots,i_{d_{I}-1},j_{0},\ldots,j_{d_{J}-1})\in n_{i_{0}}\times\ldots\times n_{i_{d_{I}-1}}\times n_{j_{0}}\times\ldots\times n_{j_{d_{J}-1}}
\end{eqnarray}
where $\cdot$ is scalar multiplication and $\pi_{A}$, $\pi_{B}$,
and $\pi_{C}$ are permutations (reorderings) of their respective
indices. These permutations are necessary to write the general definition
because dimensions may have been chosen in any position and in any
relative order when determining the labeling of indices. To simplify
the notation, the sets of index labels $i_{0}\ldots i_{d_{I}-1}$,
$j_{0}\ldots j_{d_{J}-1}$, and $p_{0}\ldots p_{d_{P}-1}$ are denoted
as \emph{index bundles} $I$, $J$, and $P$, respectively. The $I$
and $J$ bundles contain the uncontracted (free) indices, while the
$P$ bundle contains the contracted (bound) indices. Additionally,
we will make use of Einstein notation, such that the indices in the
$P$ bundle, since they appear twice on the right-hand side, are implicitly
summed over, and the remaining indices are implicitly iterated over.
Using these simplifications, the general tensor contraction becomes,
\begin{equation}
\mathscr{C}_{\pi_{C}(IJ)}\coloneqq\mathscr{A}_{\pi_{A}(IP)}\cdot\mathscr{B}_{\pi_{B}(PJ)}\label{eq:gen-contraction}
\end{equation}

Now the connection to matrix multiplication is readily apparent. Simplifying
the element-wise definition of matrix multiplication in the same way
gives,
\begin{equation}
C_{ij}=A_{ip}\cdot B_{pj}
\end{equation}
This is identical to tensor contraction except that, (1) the index
bundles $I$, $J$, and $P$ may contain more than one index, while
$i$, $j$, and $p$ are single indices, and (2), the indices in the
tensor contraction case may be arbitrarily ordered by the permutation
operators. In the matrix case, permutation of the indices in $A$,
$B$, or $C$ amounts to simple matrix transposition, but in the tensor
case indices from different bundles may be interspersed as well as
transposed overall. If each of the tensor dimensions is $\mathscr{O}(N)$,
then the tensor contraction operation requires $\mathscr{O}(N^{d_{I}+d_{J}+d_{P}})$
FLOPs (floating point operations), which is the same number of operations
as a matrix multiplication with $m=N^{d_{I}}$, $n=N^{d_{J}}$, and
$k=N^{d_{P}}$.

\section{The Traditional Approach}

In order to introduce both existing and our novel approaches to tensor
contraction, let us consider a concrete, if simple, example. Say that
we have tensors $\mathscr{A}\in\mathbb{R}^{2\times4\times3\times3}$,
$\mathscr{B}\in\mathbb{R}^{4\times4\times6}$, and $\mathscr{C}\in\mathbb{R}^{6\times3\times2\times3\times4}$,
and we wish to compute the tensor contraction,
\begin{equation}
\mathscr{C}_{abcde}\coloneqq\mathscr{A}_{cfbd}\cdot\mathscr{B}_{fea}\label{eq:ex-contraction}
\end{equation}
The tensor contraction may also be written in the form of (\ref{eq:gen-contraction})
as,
\begin{equation}
\mathscr{C}_{\pi_{C}(cdbae)}\coloneqq\mathscr{A}_{\pi_{A}(cdbf)}\cdot\mathscr{B}_{\pi_{B}(fae)}
\end{equation}
where it is clear that in this case the index bundles as defined in
the previous section are $I=cdb$, $J=ae$, and $P=f$, and the action
of the permutation operators is to reorder the indices to the order
of (\ref{eq:ex-contraction}).

The most common traditional approach to tensor contraction is to use
the similarities noted above between tensor contraction and matrix
multiplication to implement the former in terms of the latter, making
use of highly-tuned matrix multiplication routines such as through
the BLAS interface. To see how this is possible, assume for a moment
that we had picked a slightly different tensor contraction such that
$\pi_{A}$, $\pi_{B}$, and $\pi_{C}$ were the identity permutations,
\begin{equation}
\mathscr{\tilde{C}}_{cdbae}\coloneqq\mathscr{\tilde{A}}_{cdbf}\cdot\mathscr{\tilde{B}}_{fae}
\end{equation}
with the same dimension length for each index such that $\tilde{\mathscr{C}}\in\mathbb{R}^{2\times3\times3\times6\times4}$
etc.

Let us assume also that the tensors are laid out in \emph{general
column-major order}, which is similar to the well-known column-major
order for matrices. In this format, the entries of each tensor are
arranged in memory contiguously and with increasing colexicographic
order of the indices.\footnote{An example of colexicographic order for three indices is 000, 100,
200, $\ldots$, 010, 110, 210, $\ldots$, 020, $\ldots\ldots$, 001,
etc.} For example, the locations of elements of $\tilde{\mathscr{C}}$
relative to the base address are given by,
\begin{eqnarray}
loc(\tilde{\mathscr{C}}_{cdbae}) & = & c+d\cdot n_{c}+b\cdot n_{c}n_{d}+a\cdot n_{c}n_{d}n_{b}+e\cdot n_{c}n_{d}n_{b}n_{a}\nonumber \\
 & = & c+d\cdot n_{c}+b\cdot n_{c}n_{d}+(a+e\cdot n_{a})\cdot n_{c}n_{d}n_{b}\label{eq:loc}
\end{eqnarray}
where scalar multiplication of lengths is assumed. It is assumed that
the values of the indices run over their entire range in this and
similar expressions (i.e. $(a,b,c,d,e)\in n_{a}\times n_{b}\times n_{c}\times n_{d}\times n_{e}$
for (\ref{eq:loc})). Similarly, the locations of the indices in $\tilde{\mathscr{A}}$
and $\tilde{\mathscr{B}}$ are given by,
\begin{eqnarray}
loc(\mathscr{\tilde{A}}_{cdbf}) & = & c+d\cdot n_{c}+b\cdot n_{c}n_{d}+f\cdot n_{c}n_{d}n_{b}\\
loc(\mathscr{\tilde{B}}_{fae}) & = & f+a\cdot n_{f}+e\cdot n_{f}n_{a}\nonumber \\
 & = & f+(a+e\cdot n_{a})\cdot n_{f}
\end{eqnarray}
The range of values $c+d\cdot n_{c}+b\cdot n_{c}n_{d}$ for $(c,d,b)\in n_{c}\times n_{d}\times n_{b}=2\times3\times3$
is simply the ordered range of values $0\le\bar{I}<n_{\bar{I}}$ for
$n_{\bar{I}}=n_{c}n_{d}n_{b}=18$, thanks to column-major ordering.
When multiple indices may be collapsed into the range of a single
contiguous index, they are said to be \emph{sequentially} \emph{contiguous}.
Similarly the range $a+e\cdot n_{a}$ for $(a,e)\in n_{a}\times n_{e}=6\times4$
is identical to $0\le\bar{J}<n_{\bar{J}}=n_{a}n_{e}=24$, and the
possible values of $f$ are trivially the range $0\le\bar{P}<n_{\bar{P}}=n_{f}=4$.
Furthermore, since the value of the combined index $\bar{I}$ is the
same for identical values of $c$, $d$, and $b$ in both the $\mathscr{A}$
and $\mathscr{B}$ tensors, they may both be addressed by the same
\emph{linearized index }$\bar{I}$, and the same is true of $\bar{J}$
and (trivially) $\bar{P}$ as well. Thus, the tensors $\mathscr{\tilde{A}}$,
$\tilde{\mathscr{B}}$, and $\mathscr{\tilde{C}}$ are \emph{structurally
equivalent }to matrices $\tilde{A}$, $\tilde{B}$, and $\tilde{C}$
that are $n_{\bar{I}}\times n_{\bar{P}}$, $n_{\bar{P}}\times n_{\bar{J}}$,
and $n_{\bar{I}}\times n_{\bar{J}}$, respectively. The tilde denotes
that the tensor is compatible with a matrix layout and vice versa.
The tensor contraction is also \emph{functionally equivalent }(i.e.
it results in the same values in the same locations in memory) to
the matrix multiplication,
\begin{equation}
\tilde{C}_{\bar{I}\bar{J}}\coloneqq\tilde{A}_{\bar{I}\bar{P}}\cdot\tilde{B}_{\bar{P}\bar{J}}
\end{equation}
Thus, the tensor contraction may be accomplished by simply performing
a matrix multiplication with the proper parameters. This is possible
whenever the indices in each of the bundles $I$, $J$, and $P$ in
the general tensor contraction definition are sequentially contiguous
and identically ordered in each tensor (assuming general column-major
storage; the value of the linearized index must simply be the same
for identical values of the tensor indices in both tensors in the
general case)\textendash both conditions together forming the condition
of \emph{linearizabilit}y\textendash so that they may be replaced
by linearized\emph{ }indices $\bar{I}$, $\bar{J}$, and $\bar{P}$
in a matrix multiplication, where the over-bar denotes linearization.

But what about our original tensor contraction? In that case, it is
easy to see that the indices are not sequentially contiguous (for
example $b$ does not follow $d$ in $\mathscr{C}$), nor are they
identically ordered in all cases (for example, $b$ is ordered before
$c$ in $\mathscr{C}$, but $c$ is before $b$ in $\mathscr{A}$).
In this case, the tensors may be \emph{transposed} (reordered) in
memory such that the indices become sequentially contiguous and may
be linearized. Using our example, we may copy the elements in $\mathscr{A}$
to their corresponding locations in $\tilde{\mathscr{A}}$, and the
elements in $\mathscr{B}$ to $\tilde{\mathscr{B}}$, perform the
matrix multiplication, then copy the resulting elements of $\tilde{\mathscr{C}}$
to their final locations in $\mathscr{C}$. This approach is commonly
termed the TTGT (transpose-transpose-{\tt GEMM}-transpose) approach,
after the {\tt GEMM} matrix multiplication function in BLAS. In general,
if we define the tensor contraction operation as in (\ref{eq:gen-contraction})
then we may implement TTGT using the algorithm in \figref{TTDT},
assuming that the temporary tensors $\mathscr{\tilde{A}}$, $\tilde{\mathscr{B}}$,
and $\mathscr{\tilde{C}}$ are stored in general column-major order.

\begin{figure}
\hfill{}\begin{algorithmic}[1]
\Procedure{TTGT}{$\alpha,\mathscr{A}_{\pi_A(IP)},\mathscr{B}_{\pi_B(PJ)},\beta,\mathscr{C}_{\pi_C(IJ)}$}
\State Transpose $\tilde{\mathscr{A}}_{IP} \coloneqq \mathscr{A}_{\pi_A(IP)}$
\State Transpose $\tilde{\mathscr{B}}_{PJ} \coloneqq \mathscr{B}_{\pi_B(PJ)}$
\State $\tilde{C}_{\bar{I}\bar{J}} \coloneqq \alpha \sum_{\bar{P}=0}^{n_{\bar{P}}} \tilde{A}_{\bar{I}\bar{P}} \cdot \tilde{B}_{\bar{P}\bar{J}}$
\State Transpose and sum $\mathscr{C}_{\pi_C(IJ)} \coloneqq \tilde{\mathscr{C}}_{IJ} + \beta \mathscr{C}_{\pi_C(IJ)}$
\EndProcedure
\end{algorithmic}\hfill{}\caption{\label{fig:TTDT}Schematic implementation of the TTGT approach for
tensor contraction. Notational details explained in text.}

\end{figure}

This method has been used to implement tensor contraction in a vast
number of scientific applications over the past few decades, and is
the implementation used by popular tensor packages such as NumPy \cite{NumPy}
and the MATLAB Tensor Toolbox \cite{ttoolbox}. However, there are
some drawbacks to this approach:
\begin{itemize}
\item The storage space required is increased by a factor of two, since
full copies of $\mathscr{A}$, $\mathscr{B}$, and $\mathscr{C}$
are required. This storage space may either be allocated inside the
tensor contraction routine or supplied as workspace by the user, complicating
user interfaces. This deficiency is similar to that encountered in
traditional implementations of Strassen's algorithm for matrix multiplication
\cite{FLAWN79}.
\item The tensor transpositions require a (sometimes quite significant)
amount of time relative to the matrix multiplication step even after
optimization of the transposition step, as evidenced by the results
of the present work and elsewhere in the literature \cite{hartono_performance_2009,hanrath_efficient_2010,gett}.
As an applied example, after aggressively reordering operations to
reduce the number of transpositions needed in the NCC quantum chemistry
program \cite{matthews_non-orthogonal_2015} (transpositions can be
elided if the ordering required for one operation matches that required
for the next), we still measure 15-50\% of the total program time
spent in tensor transposition. This overhead is especially onerous
in methods such as CCSD(T) \cite{raghavachari_fifth-order_1989,apra_efficient_2014},
where storage size (and hence transposition cost) scales as $\mathscr{O}(N^{6})$
but computation scales only as $\mathscr{O}(N^{7})$.
\item If a comprehensive tensor contraction interface is not available (e.g.
in FORTRAN or C), or if tensor transposition must be otherwise handled
by the calling application to ensure efficiency, significant algorithmic
and code complexity is required which increases programmer burden
and the incidence of errors, while obscuring the scientific algorithms
in the application. Although it is hard to measure this deficiency
of the TTGT approach quantitatively, the long history in the literature
dealing with optimization and implementation of TTGT indicates the
level of effort implied by this approach.
\end{itemize}
These deficiencies motivate our implementation of a ``native'' (i.e.
acting directly on general tensors) high-performance tensor contraction
algorithm.

\section{New Matrix Representations of Tensors}

\subsection{The Scatter-Matrix Layout}

In order to eliminate the need for costly tensor transpositions, it
is necessary to eschew the standard BLAS interface for matrix multiplication.
However, we also wish to make use of highly-tuned algorithms for matrix
multiplication that, by the mathematical equivalence between matrix
multiplication and tensor contraction, should be applicable. The BLIS
framework gives us an opportunity to do just this. In \secref{Matrix-Multiplication},
we noted that the only operations required in the BLIS framework are
(1) matrix partitioning, (2) packing of matrix panels as slivers,
and (3) invoking the micro-kernel. The fundamental question is then,
``how can we perform these operations without changing the data layout
(transposing) the tensors?''

Going back to our example contraction, let us look at how the tensors
are laid out in the non-sequentially contiguous case and relate that
to a matrix representation. Focusing on the $\mathscr{C}$ tensor,
the indices are grouped into bundles as $I=cdb$ and $J=ae$, while
the tensor elements are laid out in general column-major order according
to the ordering $\mathscr{C}_{abcde}$. As for $\mathscr{\tilde{C}}$,
we can easily compute the locations of tensor elements in $\mathscr{C}$,
\begin{eqnarray}
loc(\mathscr{C}_{abcde}) & = & a+b\cdot n_{a}+c\cdot n_{a}n_{b}+d\cdot n_{a}n_{b}n_{c}+e\cdot n_{a}n_{b}n_{c}n_{d}\nonumber \\
 & = & (c\cdot n_{a}n_{b}+d\cdot n_{a}n_{b}n_{c}+b\cdot n_{a})+(a+e\cdot n_{a}n_{b}n_{c}n_{d})
\end{eqnarray}
where in the second equality the terms have been grouped by index
bundle. The values for either subexpression do not form a simple,
contiguous range that can be represented by a single linearized index.
However, we can compare this expression to the locations for the transposed
tensor $\tilde{\mathscr{C}}$,
\begin{eqnarray}
loc(\tilde{\mathscr{C}}_{cdbae}) & = & (c+d\cdot n_{c}+b\cdot n_{c}n_{d})+(a+e\cdot n_{a})\cdot n_{c}n_{d}n_{b}
\end{eqnarray}
and also to the locations of the matricized form $\tilde{C}$, which
we know has the same element locations as $\tilde{\mathscr{C}}$,
\begin{eqnarray}
loc(\tilde{C}_{\bar{I}\bar{J}}) & = & \bar{I}+\bar{J}\cdot n_{\bar{I}}
\end{eqnarray}

\begin{figure}
\begin{centering}
\includegraphics[width=0.6\columnwidth]{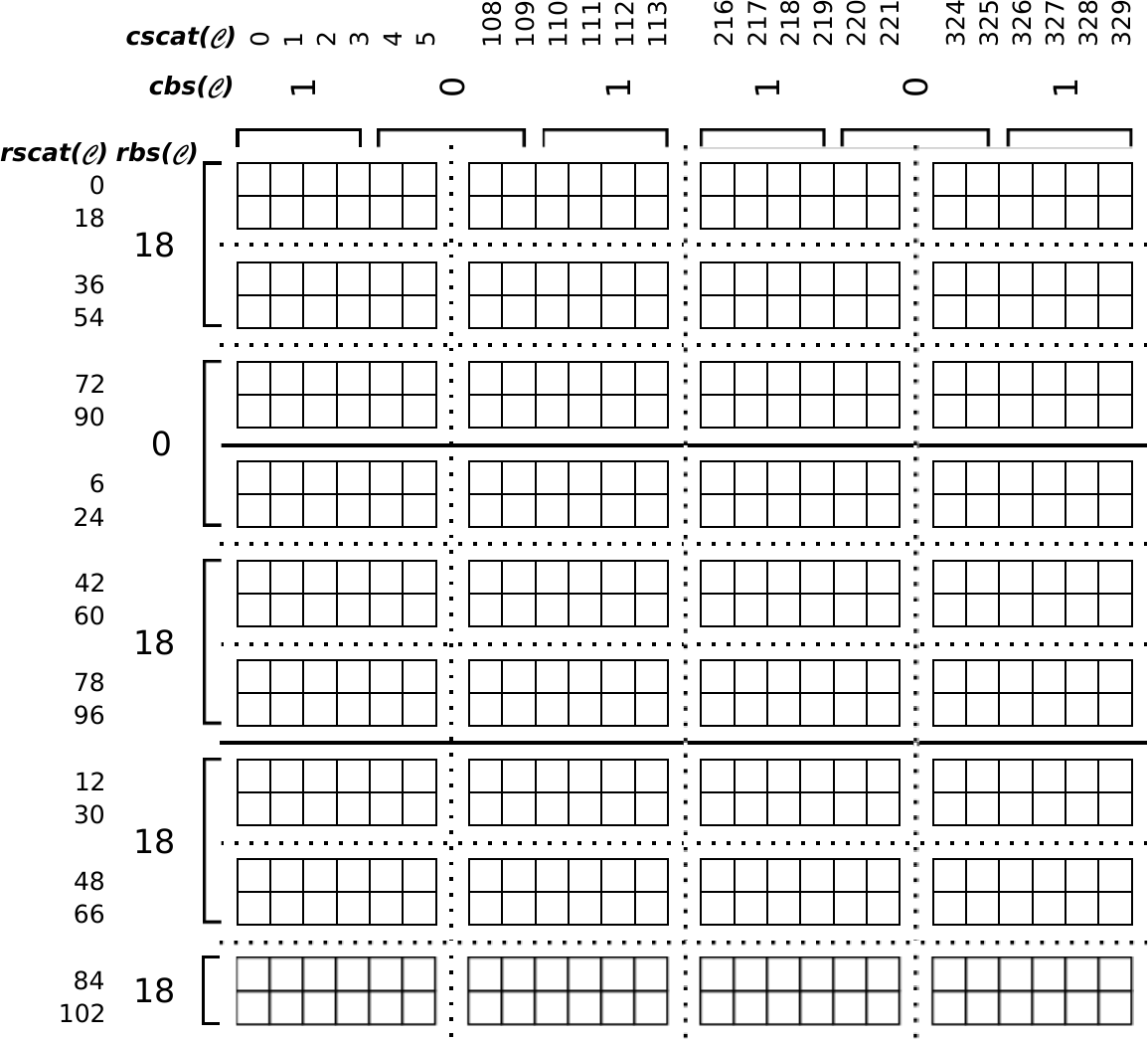}
\par\end{centering}
\caption{\label{fig:bsm}Example of a \emph{block-scatter-matrix} \emph{layout}
(see \subsecref{bsm}) for the tensor $\mathscr{C}_{abcde}\in\mathbb{R}^{6\times3\times2\times3\times4}$
with a general column-major data layout (giving strides of 1, 6, 18,
36, and 108). The matrix representation is $C_{\bar{I}\bar{J}}$ for
the bundles $I=cdb$ and $J=ae$. The blocking parameters are $m_{R}=n_{R}=4.$
Note that in this case the dimensions $c$ and $d$ are sequentially
contiguous and so a regular stride can be maintained for larger blocks.}
\end{figure}

Since $n_{\bar{I}}=n_{c}n_{d}n_{b}$, we have $\bar{I}=c+d\cdot n_{c}+b\cdot n_{c}n_{d}$
and $\bar{J}=a+e\cdot n_{a}$. Thus, given some values of $\bar{I}$
and $\bar{J}$, we can compute the values of $c$, $d$, $b$, $a$,
and $e$, and from those we can finally compute the element location
in the \emph{unmodified} $\mathscr{C}$ tensor. Performing this computation
for each individual element is likely not efficient, so the relationship
between the tensor and matrix layout may be saved in a row \emph{scatter
vector} $rscat(\mathscr{C})$, which gives the sum $c\cdot n_{a}n_{b}+d\cdot n_{a}n_{b}n_{c}+b\cdot n_{a}$
for each value of $\bar{I}$, and a column scatter vector $cscat(\mathscr{C})$,
which gives the sum $a+e\cdot n_{a}n_{b}n_{c}n_{d}$ for each value
of $\bar{J}$. For our example, we can compute these scatter vectors
as,
\begin{eqnarray}
rscat_{\bar{I}}(\mathscr{C}) & = & c\cdot n_{a}n_{b}+d\cdot n_{a}n_{b}n_{c}+b\cdot n_{a}\nonumber \\
 & = & \left(\bar{I}\mod n_{c}\right)\cdot n_{a}n_{b}+\left(\frac{\bar{I}}{n_{c}}\mod n_{d}\right)\cdot n_{a}n_{b}n_{c}\nonumber \\
 &  & +\left(\frac{\bar{I}}{n_{c}n_{d}}\mod n_{b}\right)\cdot n_{a}\\
cscat_{\bar{J}}(\mathscr{C}) & = & a+e\cdot n_{a}n_{b}n_{c}n_{d}\nonumber \\
 & = & \left(\bar{J}\mod n_{a}\right)+\left(\frac{\bar{J}}{n_{a}}\mod n_{e}\right)\cdot n_{a}n_{b}n_{c}n_{d}\label{eq:cscat-c}
\end{eqnarray}
The general location of elements in an $n_{\bar{I}}\times n_{\bar{J}}$
matrix $C$ may then be given in the tensor layout of $\mathscr{C}$
as,
\begin{equation}
loc(C_{\bar{I}\bar{J}})=rscat_{\bar{I}}(\mathscr{C})+cscat_{\bar{J}}(\mathscr{C})
\end{equation}
We call this additive formula for the locations of tensor elements
mapped to matrix elements and the associated scatter vectors the \emph{scatter-matrix
layout}, in which the elements of a tensor stored in its natural physical
layout may be addressed as if they were a matrix. For our example
$\mathscr{C}$, the values of the scatter vectors are illustrated
in \figref{bsm}. For tensor contraction, the scatter vectors $rscat(\mathscr{A})$
and $rscat(\mathscr{C})$ are defined for each value of $\bar{I}$,
while the scatter vectors $cscat(\mathscr{B})$ and $cscat(\mathscr{C})$
are defined for each value of $\bar{J}$, and $cscat(\mathscr{A})$
and $rscat(\mathscr{B})$ for each value of $\bar{P}$.

In general, a tensor may be stored in general column-major order,
but also in the similar general row-major order or in any other layout
that preserves the uniqueness of elements. The location of elements
in a generic layout for $\mathscr{C}$ may be described by a set of
index \emph{strides}, $s(\mathscr{C})$. The location of elements
in $\mathscr{C}$ is then,
\begin{equation}
loc(\mathscr{C}_{IJ})=i_{0}\cdot s_{i_{0}}(\mathscr{C})+\ldots+i_{d_{I}-1}\cdot s_{i_{d_{I}-1}}(\mathscr{C})+j_{0}\cdot s_{j_{0}}(\mathscr{C})+\ldots+j_{d_{J}-1}\cdot s_{j_{d_{J}-1}}(\mathscr{C})
\end{equation}
Unlike the dimension lengths, the strides must always be denoted in
reference to a particular tensor because their value for indices which
are labeled the same in multiple tensors does not need to agree for
each tensor. For our example $\mathscr{C}$ we have $(s_{a}(\mathscr{C}),s_{b}(\mathscr{C}),s_{c}(\mathscr{C}),s_{d}(\mathscr{C}),s_{e}(\mathscr{C}))=(1,6,18,36,108)$.
The scatter vectors for a generic layout may also be simply computed,
\begin{equation}
rscat_{\bar{I}}(\mathscr{A})=i_{0}\cdot s_{i_{0}}(\mathscr{C})+\ldots+i_{d_{I}-1}\cdot s_{i_{d_{I}-1}}(C)
\end{equation}
and similarly for the other scatter vectors.

Turning to the operations necessary to implement the BLIS framework,
we can see that the first, matrix partitioning, is trivially implemented
by partitioning of the row and/or scatter vectors. For example, the
top-right quadrant of a matrix in the scatter-matrix layout may be
represented by a sub-matrix with the first half of the full row scatter
vector and the second half of the full column scatter vector. The
$A$, $B$, and $C$ matrices (matrix representations of $\mathscr{A}$,
$\mathscr{B}$, and $\mathscr{C}$ in the scatter-matrix layout) may
then be partitioned as much as is necessary while maintaining their
scatter-matrix layouts, with essentially no overhead.

For the packing operations, the usual kernels as implemented in BLIS
cannot be used, because they assume a general matrix layout, with
a constant row and column stride (as in the general tensor layout
described above). The scatter-matrix layout, though, does not have
a constant stride as can be seen in \figref{bsm}. A new packing kernel
is required that takes row and column scatter vectors and loads elements
accordingly. We have implemented such a packing kernel for general
scatter-matrix layouts. This packing kernel is not as efficient as
the ``normal'' matrix packing kernel because the scatter-matrix
layout inhibits vectorized loads of elements and requires a higher
memory bandwidth due to the need to read the scatter vector entries.

Finally, when invoking the micro-kernel, the scatter-matrix layout
cannot be used directly. Because the micro-kernel is hand-written
in assembly code, and assumes constant row and column strides for
$C$, the micro-kernel must be invoked to write to a small temporary
buffer (of size $m_{R}\times n_{R}$), and then the values from this
buffer written out to memory according the the scatter-matrix layout
of $C$. This is again a source of inefficiency as the micro-kernel
is normally tuned to write out the elements of $C$ using vector writes.
The micro-kernel does not need to be modified for scatter-matrix layouts
of $A$ and $B$ because the packing operations write the values in
the buffers $\tilde{A}_{i}$ and $\tilde{B}_{p}$ in a fixed layout
independent of the input layout.

The implementation of tensor contraction using the BLIS framework
augmented by scatter-matrix layouts for $A$, $B$, and $C$ with
concordant packing kernels and micro-kernel adjustments is termed
the Scatter-Matrix Tensor Contraction (SMTC) algorithm. The next section
extends the scatter-matrix layout and SMTC to avoid most of the inefficiency
incurred by the use of scatter vectors in the packing and micro-kernel
operations.

\subsection{The Block-Scatter-Matrix Layout\label{subsec:bsm}}

Since the micro-kernel is the basic unit of work in BLIS, the size
of a micro-kernel update defines a natural blocking of the matrix
dimensions $m$ and $n$. These block sizes are denoted $m_{R}$ and
$n_{R}$ and are usually 4, 6, or 8 for double-precision real numbers.
So, while the values of the scatter vectors $rscat(\mathscr{C})$
and $cscat(\mathscr{C})$, are in general not ``well-behaved'' (monotonically
increasing, regularly spaced, etc.) over their entire range which
would allow a regular scalar stride to be used instead, they may often
be well-behaved for short stretches.

For example, $cscat(\mathscr{C})$ for our sample tensor contraction
as calculated from (\ref{eq:cscat-c}) begins with the sequence $0,1,\ldots,n_{a}-1=5$,
then jumps up to $n_{a}n_{b}n_{c}n_{d}=108$, continues $108,109,\ldots108+n_{a}-1$=113,
jumps again and so on. So, for small stretches that don't cross one
of the large jumps, $C_{\bar{I}\bar{J}}$ behaves very much like a
row-major matrix with a unit ``stride'' for $\bar{J}$. Similarly,
for small stretches of $\bar{I}$, $rscat(\mathscr{C})$ also behaves
as if it were a constant stride of $n_{a}n_{b}=18$.

Thus, when $C_{\bar{I}\bar{J}}$ is partitioned into $m_{R}\times n_{R}$
micro-tiles, it is quite possible that the ranges of $\bar{I}$ and
$\bar{J}$ for many of these micro-tiles may fall entirely into one
of these constant stride regions. In fact, the fraction of micro-tiles
that do so is 53\% for our example, and generally a much higher fraction
for real problems with larger tensor dimensions. When a micro-tile
of $C_{\bar{I}\bar{J}}$ has constant strides then it may be fed directly
into the micro-kernel without writing to a temporary buffer first.
To make use of this optimization, a value is stored for each $m_{R}$-length
block of $rscat(\mathscr{C})$ and each $n_{R}$-length block of $cscat(\mathscr{C})$
which is either the stride for this block, when it falls in a region
of constant stride, or the value 0 to denote that no such stride exists.
These values are stored in row and column \emph{block-scatter vectors
}$rbs(\mathscr{C})$ and $cbs(\mathscr{C})$, as illustrated in \figref{bsm}.

For a general row or column scatter vector $scat(\mathscr{T})$ of
length $l$, we may use a blocking parameter $b$ to create a block-scatter
vector $bs(\mathscr{T})$ of length $\left\lceil \frac{l}{b}\right\rceil $.
The $i$th entry of the block-scatter vector $bs_{i}(\mathscr{T})$
is equal to some positive integer $s$ if $scat_{j}(\mathscr{T})-scat_{j-1}(\mathscr{T})=s$
for all $i\cdot b<j<min(l,(i+1)\cdot b)$ and equal to 0 otherwise.

The row scatter vector $rscat(\mathscr{A})$ and the column scatter
vector $cscat(\mathscr{B})$ may be blocked into block-scatter vectors
by $m_{R}$ and $n_{R}$, respectively, but the scatter vectors for
the $P$ bundle ($cscat(\mathscr{A})$ and $rscat(\mathscr{B})$)
do not have any natural blocking smaller than $k_{C}$ in the BLIS
approach. Since $k_{C}$ is usually quite large (e.g. 256 on the Intel
Haswell architecture for double precision), it is unlikely that generating
block-scatter vectors with this block size will yield any benefit.
Instead, we introduce an additional blocking parameter $k_{P}$ that
is small (on the order of $m_{R}$ and $n_{R}$) with which to generate
block-scatter vectors $cbs(\mathscr{A})$ and $rbs(\mathscr{B})$.
To make use of this block-scatter vector, the packing operation on
a $m_{R}\times k_{C}$ sliver of $\mathscr{A}$ or a $k_{C}\times n_{R}$
sliver of $\mathscr{B}$ is broken down into a sequence of $m_{R}\times k_{P}$
or $k_{P}\times n_{R}$ micro-tile packing operations, which take
advantage of constant strides from the block-scatter vectors whenever
possible.

The implementation of tensor contraction using the BLIS framework
augmented by block-scatter-matrix layouts for $A$, $B$, and $C$
with concordant packing kernels and micro-kernel adjustments is termed
the Block-Scatter-Matrix Tensor Contraction (BSMTC) algorithm. While
we have implemented both SMTC and BSMTC, the clear advantages of BSMTC
over SMTC leads us to report performance results only for the former.

\subsection{Using the (Block-)Scatter-Matrix Layout}

The scatter-matrix and block-scatter-matrix layouts require storage
of the scatter and block scatter vectors for each of the index bundles.
We wish to avoid allocating these vectors at the start of the tensor
contraction algorithm for several reasons,
\begin{itemize}
\item This requires at least one general memory allocation ({\tt malloc})
per call which may involve virtual memory operations (e.g. {\tt mmap}).
\item This requires unbounded (i.e. scaling with input tensor size) additional
storage ($\mathscr{O}(n_{\bar{I}}+n_{\bar{J}}+n_{\bar{P}})$, although
the total amount is still much less than the $\mathscr{O}(n_{\bar{I}}n_{\bar{J}}+n_{\bar{I}}n_{\bar{P}}+n_{\bar{P}}n_{\bar{J}})$
required in TTGT).
\item This complicates the interface because users may want to supply external
workspace or precomputed scatter vectors, etc.
\end{itemize}
Instead, we delay the transition to a scatter- or block-scatter-matrix
layout until the input and output matrices have been partitioned into
panels of fixed (bounded) size. For the tensors $\mathscr{A}$ and
\textbf{$\mathscr{B}$ }this occurs when panels $\tilde{A}_{i}$ and
$\tilde{B}_{p}$ are packed into contiguous storage (see \figref{blis}),
and for $\mathscr{C}$ this occurs just before entry into the inner
kernel (the last two loops around the BLIS micro-kernel). At this
stage, the maximum size of the scatter vectors is known, and they
can be allocated from persistent storage in the same manner as the
pack buffers $\tilde{A}_{i}$ and $\tilde{B}_{p}$. In our implementation,
the size of the memory allocations for the pack buffers is increased
slightly to accommodate the scatter vectors for $\mathscr{A}$ and
$\mathscr{B}$, and a separate buffer is only required for the scatter
vectors of $\mathscr{C}$.

We then have four layout types which are encountered during the contraction
algorithm: general tensor layout, scatter-matrix layout, block-scatter-matrix
layout, and packed matrix layout (for $\tilde{A}_{i}$ and $\tilde{B}_{p}$).
The way these layout types are handled in the key operations in the
BLIS approach are summarized in \figref{layouts}, along with a comparison
to a normal matrix layout (which is handled essentially the same as
the packed matrix layout).

\begin{figure}
\begin{centering}
\begin{tabular}{|>{\centering}p{0.1\columnwidth}||>{\centering}p{0.26\columnwidth}|>{\centering}p{0.22\columnwidth}|>{\centering}p{0.22\columnwidth}|}
\hline 
{\footnotesize{} Layout Type} & {\footnotesize{}When Partitioning} & {\footnotesize{}When Packing} & {\footnotesize{}After each Micro-kernel Invocation}\tabularnewline
\hline 
\hline 
{\footnotesize{}(Packed) Matrix} & {\footnotesize{}Keep track of $u$ and $v$ implicitly by adjusting
base pointer.} & {\footnotesize{}Reference elements using base pointer and matrix row
and column strides.} & {\footnotesize{}Update to $M_{uv}$ done in micro-kernel.}\tabularnewline
\hline 
{\footnotesize{}Scatter-Matrix} & {\footnotesize{}Keep track of offset in $rscat(\mathscr{T})$ and
$cscat(\mathscr{T})$ vectors (no change to base pointer).} & {\footnotesize{}Reference elements using base pointer and $rscat(\mathscr{T})$,
$cscat(\mathscr{T})$.} & {\footnotesize{}Accumulate from micro-kernel into buffer, then scatter
into $T_{\bar{I}\bar{J}}$.}\tabularnewline
\hline 
{\footnotesize{}Block-Scatter-Matrix} & {\footnotesize{}Keep track of offset in $\{r,c\}scat(\mathscr{T})$
and $\{r,c\}bs(\mathscr{T})$ vectors. If $rbs(\mathscr{T})$ and/or
$cbs(\mathscr{T})$ are constant stride for the current block, adjust
base pointer.} & {\footnotesize{}Pack as regular matrix if $rbs(\mathscr{T})$ and
$cbs(\mathscr{T})$ for the current block are valid, as a scatter-matrix
otherwise.} & {\footnotesize{}Use micro-kernel update if $rbs(\mathscr{T})$ and
$cbs(\mathscr{T})$ are valid. Treat as scatter-matrix otherwise.}\tabularnewline
\hline 
{\footnotesize{}Tensor} & {\footnotesize{}Keep track of the current values of $\bar{U}$ and
$\bar{V}$ and compute $\{u_{k}\}_{k=0}^{d_{U}-1}$ etc. when necessary.} & {\footnotesize{}n/a} & {\footnotesize{}n/a}\tabularnewline
\hline 
\end{tabular}
\par\end{centering}
\caption{\label{fig:layouts}Handling of tensor layout types in important BLIS
kernels. Each layout is assumed to refer to a tensor $\mathscr{T}_{\pi_{T}(UV)}$
(one of $\mathscr{A}$, $\mathscr{B}$, or $\mathscr{C}$) for some
index bundles $U$ and $V$ and its matrix representation $T_{\bar{U}\bar{V}}$,
while matrix layouts refer to a general matrix $M_{uv}$. Note that
packing of block-scatter-matrix layouts may also take advantage of
cases where only one of $rbs(\mathscr{T})$ and $cbs(\mathscr{T})$
indicates a constant stride.}

\end{figure}

\section{Implementation Details}

\subsection{Framework Implementation}

The SMTC and BSMTC algorithms we have implemented use a BLIS-like
framework and the actual BLIS micro-kernels. The reasons we did not
pursue using the BLIS framework directly are subtle. On one hand,
the BLIS framework does offer substantial flexibility in defining
custom operations such as packing kernels, but not quite the level
of flexibility we require especially with regards to the loops inside
the macro-kernel and at the level of the micro-kernel. On the other
hand, we also wished to explore alternative techniques for implementing
the BLIS approach. One significant difference of our implementation
to BLIS is that where BLIS uses a recursive data structure to specify
the necessary partitioning and packing operations at runtime (the
\emph{control tree}), we use a C++11 variadic template structure which
enables the compiler to automatically select the appropriate partitioning
and packing operations based on the type of the tensor or matrix object
passed to each step.

For example, all operands begin as {\tt Tensor<T>} objects which
are straight-forward C++ objects representing general tensors, but
are then wrapped in {\tt TensorMatrix<T>} objects that divide the
indices into bundles and keep track of partitioning of the matrix
representation of the tensor. These objects are eventually ``matrified''
into {\tt BlockScatterMatrix<T>} objects by generating scatter and
block scatter vectors for the current matrix partition. Finally, $A_{\bar{I}\bar{P}}$
and $B_{\bar{P}\bar{J}}$ are packed into regular {\tt Matrix<T>}
objects using the block-scatter matrix layout. Overloaded packing
and micro-kernel wrapper functions handle each of these types as appropriate.
Our implementation does use the various assembly micro-kernels and
blocking parameters from BLIS, though, and in testing we have found
no measurable difference in performance of matrix multiplication.
The C++ template specification for the BSMTC driver is given in \figref{bsmtc}.

\begin{figure}
\hfill{}
\begin{lstlisting}[basicstyle={\normalsize\ttfamily}]
// TensorMatrix<T> A, B, C;

GEMM<PartitionN<NC>,
     PartitionK<KC>,
     MatrifyAndPackB<KP,NR>,
     PartitionM<MC>,
     MatrifyAndPackA<MR,KP>,
     MatrifyC<MR,NR>,
     PartitionN<NR>,
     PartitionM<MR>,
     MicroKernel<MR,NR>
    >::run<T> gemm;

// comm is the thread communicator
// (threading details not shown)
gemm(comm, alpha, A, B, beta, C);
\end{lstlisting}
\hfill{}

\caption{\label{fig:bsmtc}Variadic template implementation of BSMTC. The steps
specified in the {\tt GEMM<...>} template can be directly compared
to those in \figref{blis} with the addition of tensor ``matrification''
(conversion from {\tt TensorMatrix<T>} to {\tt BlockScatterMatrix<T>}).}

\end{figure}

In this template implementation, the data type (float, double, etc.)
is a template parameter, but the number of dimensions in each tensor
is not. We have found requiring this parameter to be a compile-time
constant to be overly restrictive in practice. However, because the
dimensionality is not known \emph{a priori} (or bounded), the interface
layer must perform some small memory allocations to manipulate arrays
of dimension lengths, strides, labels, etc. Most {\tt malloc} implementations
contain optimizations for very small allocations however (for example,
Apple's OS X has a per-thread pool for blocks of up to 992 bytes),
and indeed we do not measure a significant overhead in our implementation
due to these allocations. Additionally, dimension labels (represented
by a string type) almost always fall in the range of the small string
optimization (SSO) which eliminates some allocations. If handling
short vectors becomes an issue, it is also feasible to enforce a maximum
dimensionality for tensors so that a static allocation or the stack
may be used.

\subsection{Multithreading}

Run-time information controlling the tensor contraction operation
is stored in the variadic template object on a per-thread basis. This
information includes the address of the packing buffers and the desired
level of parallelism at each loop, but also information about the
current \emph{thread communicator}. The thread communicator is a concept
that was adopted in BLIS to aid in parallelization \cite{blis-multi,blis2},
and closely resembles the communicator concept from MPI. Each communicator
references a shared barrier object which the threads utilize for synchronization.
Parallelism is managed at each level by splitting the communicator
into a set of sub-communicators to which work is assigned. As in BLIS,
there are five parallelizable loops as can be seen in \figref{blis},
although the loop over dimension $p$ (with block size $k_{C}$) is
not parallelized since this would require additional synchronization
and/or temporary buffers with reduction.

\subsection{Tensor Layout and Access}

For both matrices and tensors, it is important to attempt to access
elements in a way that maximizes spatial locality of the data. For
matrices where one of the strides is unit (as is always true in the
BLAS interface), this means simply ordering loops such that the stride-1
(unit stride) dimension is iterated over in the inner loop. This affects
access during packing and during the update to $C\coloneqq AB$ in
the micro-kernel. Since the micro-kernel is usually hand-written in
assembly, it is simpler to hard-code a preference for unit row stride
and to compute the equivalent operation $C^{T}\coloneqq B^{T}A^{T}$
in the case that the column stride is unit instead. This assures high
performance for all eight variants of matrix multiplication depending
on transposition of each operand. For tensors, however, the number
of possible types and combinations of transpositions is enormous.
Additionally, it may not be possible to simultaneously guarantee stride-1
access in all operands if dimensions appear in different orders in
two tensors. For example, in the contraction $\mathscr{C}_{ijk}\coloneqq\mathscr{A}_{jli}\mathscr{B}_{lk}$
in general column-major layout it is impossible to achieve stride-1
access in both $\mathscr{A}$ and $\mathscr{C}$ simultaneously.

Because of this complication, we heuristically reorder the tensor
dimensions within each bundle $I$, $J$, and $P$ from their original,
user-defined ordering. Since this reordering is purely logical, it
does not require movement of the tensor data as in tensor transposition,
and only affects the order in which tensor dimensions are iterated
over during the contraction algorithm. Additionally, while the dimensions
are being processed, any dimensions of length 1 may be removed since
the stride along these dimensions is meaningless. Lastly, dimensions
that are sequentially contiguous in all cases may be folded into a
single dimension, which decreases the indexing overhead and increases
the number of regular stride blocks in the scatter vectors. The heuristic
steps employed are:
\begin{enumerate}
\item Remove any dimensions of length one.
\item Fold dimensions in $I$, $J$, and/or $P$ that are sequentially contiguous
in all tensors.
\item Sort the dimensions in $I$ and $J$ by increasing stride in $\mathscr{C}$.
\item If $s_{j_{0}}(\mathscr{C})=1$, swap $\mathscr{A}$ with $\mathscr{B}$
and $I$ with $J$.
\item Sort the dimensions in $P$ by increasing stride in $\mathscr{A}$.
\end{enumerate}
The optimal ordering of dimensions to reduce the number of cache and
TLB misses may differ from that achieved by the above heuristics,
but these steps at least ensure that $C_{\bar{I}\bar{J}}$ has unit
row stride if possible, enabling efficient updating in the micro-kernel,
and ensure that $\mathscr{A}$ has priority for stride-1 access over
$\mathscr{B}$ since it is packed more frequently. 

The data layout and ordering of dimensions specified by the user can
have a significant impact on performance for tensor contraction, since
certain orderings may prohibit stride-1 access regardless of logical
reordering. As for matrices, ensuring that at least one stride in
each tensor is unit is the simplest condition that the user can check
for performance. However, there are several other guidelines for tensor
layout which can aid in maximizing tensor contraction efficiency:
\begin{itemize}
\item Dimensions should be ordered the same in each tensor (or more generally,
dimensions should have strides in each tensor which are ordered the
same in magnitude). 
\item Dimensions should be sequentially contiguous where possible. Alignment
of the first leading dimension (the stride of the second dimension)
may be influential on some architectures.
\item The dimensions of largest size should have the shortest strides, and
the dimension with stride 1 should especially be as long as possible.
\end{itemize}
Related work on explicit tensor transpositions (see for example \cite{ttc})
may also provide a systematic way to optimize the ordering of tensor
dimensions and help overcome inefficiencies in tensor layout.

\section{Related Work\label{sec:Related-Work}}

As mentioned previously, there are a variety of tensor-related packages
available for popular programming platforms such as NumPy \cite{NumPy}
for Python, the Tensor Toolbox \cite{ttoolbox} in MATLAB, the template
libraries Eigen \cite{Eigen} and Blitz++ \cite{Blitz} in C++, among
many others. These libraries provide a simple and intuitive interface
for creating and manipulating tensors, for example providing traditional
array-style access to individual elements, managing transposition
and reshaping, etc. Tensor contraction facilities are provided in
many of these libraries, either using explicit (although sometimes
compiler-generated) loop-based code, or using the TTGT approach. In
some applications a non-high performance library is sufficient, but
the deficiencies of the TTGT approach are highly relevant for high-performance
code. Additionally, the quality of and interface provided for tensor
operations varies significantly from library to library. It is our
hope that the high-performance and self-contained (since it does not
require large amounts of workspace) implementation provided by BSMTC
can provide a standard level of performance and functionality.

Within specific scientific domains, high-performance libraries have
appeared that implement the TTGT approach. For example, in quantum
chemistry there are software packages such as the Tensor Contraction
Engine (TCE) \cite{tce}, Cyclops Tensor Framework \cite{ctf}, libtensor
\cite{libtensor}, and TiledArray \cite{tiledarray1,tiledarray2}
that provide general tensor and in some cases specific quantum chemistry-related
functionality. Many of these libraries could benefit directly from
a native tensor contraction kernel since they focus primarily on distributed-memory
algorithms and tensor blocking for algorithmic and space efficiency.
Other approaches such as Direct Product Decomposition (DPD) packing
\cite{DPD} are specifically focused on improving the efficiency of
the TTGT approach, but could also be used on top of the BSMTC algorithm.

Other research has focused on improving the efficiency of the TTGT
approach through optimization of the tensor transposition step (and
other associated operations in quantum chemistry). Explicit searches
of the space of tensor transpose algorithms along with code generation
techniques has been used to generate high-performance tensor transpose
kernels \cite{ttc}. Tensor transposition along with handling of index
permutation symmetry in the TTGT approach has been addressed specifically
in the chemistry community \cite{hartono_performance_2009,hanrath_efficient_2010,lyakh_efficient_2015}.

One alternative to TTGT not previously discussed is the use of tensor
slicing. In this approach, the dimensionality of each tensor in the
contraction operation is successively reduced by explicitly looping
over lower-dimensional tensor contraction sub-problems. When enough
dimensions have been eliminated in this way, the inner kernel becomes
one of the standard BLAS operations, although depending on which dimensions
have been eliminated the inner kernel may be a level 2 (matrix-vector)
or even level 1 (vector-vector) operation rather than matrix multiplication.
Analysis of the resulting inner kernel can be used to optimally eliminate
indices to produce an efficient algorithm \cite{slice1,slice2}, while
for certain contraction types performance modeling and auto-tuning
have been used to generate efficient parallel implementations \cite{slice3}.
Tensor slicing has also been applied to tensor contraction on GPUs
\cite{slice_gpu1}. These approaches can also be considered native
tensor contraction algorithms since they do not require explicit tensor
transposition and may offer an alternative path to high-performance
implementations; however we do not directly compare to tensor slicing
since no standard algorithm or library has emerged for this approach,
and also because in practice tensor slicing is restricted to those
tensor contractions which allow for appropriate stride-1 access in
the matrix multiplication kernel. Additionally, since virtually all
approaches to tensor slicing in the literature rely on code generation
techniques that are specific to the number of tensor dimensions, particular
tensor contraction desired, and in some cases dimension lengths, they
cannot be directly equated to the TTGT approach and to our work, where
any tensor contraction may be computed regardless of dimensionality,
shape, and size. For a comparison of tensor slicing compared to TTGT
and other code-generation approaches to tensor contraction, see \cite{gett}.

Lastly, another native tensor contraction approach has recently been
developed independently by Springer and Bientinesi \cite{gett}, termed
GETT. The GETT algorithm is similar to BSMTC in that elements from
the input tensors are packed into fixed-size buffers to improve cache
reuse, and that a small micro-kernel is used at the basic unit of
work. However, there are several critical differences between GETT
and BSMTC. Firstly, BSMTC represents tensors as a special matrix layout,
which transparently allows any type of logical matrix operation (e.g.
partitioning) to be performed with no changes, while GETT preserves
the full-dimensional tensor structure throughout the computation.
In practice, this may limit the flexibility of GETT with respect to
the choice of micro-kernel size and cache blocking parameters since
these parameters must evenly divide the lengths of the tensor dimensions
to which they correspond, while on the other hand this eliminates
the edge cases requiring the full scatter vector in BSMTC. Secondly,
GETT uses a generated micro-kernel and heuristically determined cache
blocking parameters, which may not reach the same level of efficiency
as the assembly-coded micro-kernel and fine-tuned parameters in BLIS.
However, GETT can also adapt to varying tensor shapes while the BLIS
parameters are fixed. Lastly, BSMTC is an entirely run-time algorithm
that can operate on any size or shape of tensor, while GETT uses heuristically
guided search and code generation to implement tensor contractions
for a fixed size and shape. We have adopted the Tensor Contraction
Benchmark \cite{tensor_benchmark} from \cite{gett} for some of the
results presented in this work, as the benchmark spans a variety of
literature-derived tensor contractions from fields such as quantum
chemistry. The results using this benchmark presented here are roughly
comparable to those in \cite{gett}.

\section{Results\label{sec:Results}}

\subsection{Experimental Setup}

All of the experiments performed were run in double precision on a
single Intel Xeon E5-2690 v3 processor running at 2.6+ GHz on the
Lonestar 5 system at the Texas Advanced Computing Center. The BLIS
{\tt haswell} micro-kernel exploits the AVX2 and FMA3 features of
the Haswell architecture supported by this chip. The processor has
twelve cores with private 32KB L1 data and 256KB L2 caches, while
the 30MB L3 cache is shared among all twelve cores. The theoretical
peak floating point performance of this processor is 41.6 to 56 GFLOPs
(billion floating point operations per second) depending on the clock
boost from Intel's Turbo Boost. The practical peak performance, as
measured by timing the Intel Math Kernel Library (MKL) on large matrix
multiplications is $\sim45$ GFLOPs on one core and $\sim500$ GFLOPs
on 12 cores ($\sim41.7$ GFLOPs/core). The blocking parameters used
were: $m_{R}=6$, $n_{R}=8$, $k_{P}=4$, $m_{C}=72$, $n_{C}=4080$,
and $k_{C}=256$ for double-precision elements, which is consistent
with those used in BLIS. We compiled the BLIS micro-kernel and our
own framework with the Intel Composer XE 2016 Update 1 compilers.
Each experiment was run a number of times to warm the caches, and
the run with the lowest time (highest performance) is reported. The
multithreaded experiments used the Intel OpenMP runtime with threads
pinned to their respective cores (single-threaded runs were pinned
to an arbitrary core). All experiments were performed on double-precision
data types. The unmodified BLIS library was used for all matrix multiplications,
since it is directly comparable to our tensor contraction algorithms.
BLIS was shown to achieve performance within a few percentage points
of widely-used alternatives such as OpenBLAS \cite{openblas1,openblas2}
and within 10\% of MKL on similar processors \cite{blis1,blis2}.
The TTGT approach was implemented in our benchmarks by the TTT algorithm
of the MATLAB Tensor Toolbox v2.6 \cite{ttoolbox} in MATLAB R2016b,
with an additional call to the MATLAB {\tt permute} function to return
the $\mathscr{C}$ tensor to the proper layout (the layout of the
output $\mathscr{C}$ in the TTT algorithm is always the matricized
form $\tilde{\mathscr{C}}_{IJ}$).

In each performance graph, the $y$-axis runs from zero to the theoretical
peak performance for one core of the processor, expressed in GFLOPs
or GFLOPs/core for multi-core results, running at the maximum Turbo
Boost frequency. Since the Turbo Boost feature of newer Intel processors
can produce highly dynamic performance properties, we attempted to
control for this effect by running all of the experiments in our suite
twice, and then reporting the numbers from the second run. This process
should cause the processor to heat up sufficiently to bring the frequency
down to a stable value.

\subsection{Randomly Generated Tensor Contractions}

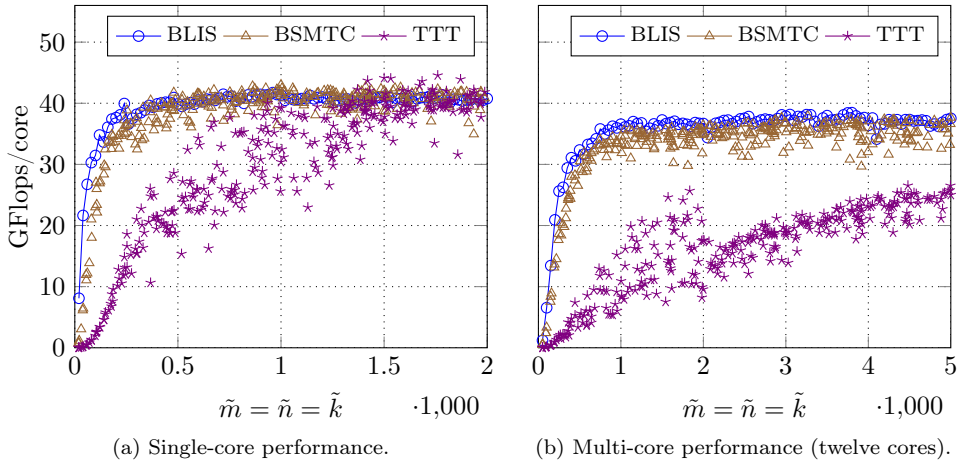
\begin{figure}
\subfloat[Single-core performance.]{\thispagestyle{empty}
\begin{tikzpicture}
\pgfplotsset {grid style={dotted,color=black}}
\begin{axis}[
scale=0.8,
grid=major,
ymin=0,
ymax=56,
ylabel={GFlops/core},
ylabel shift=-5,
ytick distance=10,
xmin=0,
xmax=2000,
scaled x ticks=real:1000,
xlabel={$\tilde m=\tilde n=\tilde k$},
legend style={legend pos=north east,font=\footnotesize},
legend columns=3
]
\addplot [mark=o,color=blue] coordinates {(20,8.097267e+00) (40,2.164350e+01) (60,2.673755e+01) (80,3.029578e+01) (100,3.145197e+01) (120,3.480363e+01) (140,3.365095e+01) (160,3.607856e+01) (180,3.741112e+01) (200,3.765175e+01) (220,3.809953e+01) (240,3.995162e+01) (260,3.659600e+01) (280,3.754494e+01) (300,3.818948e+01) (320,3.845028e+01) (340,3.888563e+01) (360,3.963436e+01) (380,3.934869e+01) (400,3.989817e+01) (420,4.006577e+01) (440,4.017178e+01) (460,4.015703e+01) (480,4.092949e+01) (500,3.990098e+01) (520,3.948734e+01) (540,3.986113e+01) (560,4.008697e+01) (580,3.985005e+01) (600,4.052882e+01) (620,4.024877e+01) (640,4.067375e+01) (660,4.063356e+01) (680,4.046237e+01) (700,4.082141e+01) (720,4.147191e+01) (740,4.106427e+01) (760,4.117178e+01) (780,4.034246e+01) (800,4.081552e+01) (820,3.998155e+01) (840,4.135895e+01) (860,4.107276e+01) (880,4.136557e+01) (900,4.144803e+01) (920,4.132290e+01) (940,4.141578e+01) (960,4.173869e+01) (980,4.143144e+01) (1000,4.133956e+01) (1020,4.048889e+01) (1040,4.082229e+01) (1060,4.059241e+01) (1080,4.066226e+01) (1100,4.075868e+01) (1120,4.075688e+01) (1140,4.127602e+01) (1160,4.072762e+01) (1180,4.093667e+01) (1200,4.101429e+01) (1220,4.075910e+01) (1240,4.104393e+01) (1260,4.045743e+01) (1280,4.042048e+01) (1300,4.066358e+01) (1320,4.071082e+01) (1340,4.043828e+01) (1360,4.006861e+01) (1380,4.043066e+01) (1400,4.070760e+01) (1420,4.068226e+01) (1440,4.101966e+01) (1460,4.084424e+01) (1480,4.098205e+01) (1500,4.113865e+01) (1520,4.106898e+01) (1540,4.012872e+01) (1560,4.081506e+01) (1580,4.070816e+01) (1600,4.083356e+01) (1620,4.120324e+01) (1640,4.024320e+01) (1660,4.072736e+01) (1680,4.092461e+01) (1700,4.091767e+01) (1720,4.113418e+01) (1740,4.077779e+01) (1760,4.095775e+01) (1780,4.137981e+01) (1800,4.043600e+01) (1820,4.015267e+01) (1840,4.066780e+01) (1860,4.057490e+01) (1880,4.036986e+01) (1900,4.051674e+01) (1920,4.055246e+01) (1940,4.090739e+01) (1960,4.052524e+01) (1980,4.050277e+01) (2000,4.077420e+01)};
\addplot [mark=triangle,draw=none,color=brown!80!black] coordinates {(18.3154,7.038611e-01) (21.2532,1.163004e+00) (18.5664,8.423279e-01) (30.7796,3.048614e+00) (38.6196,6.151860e+00) (40.6559,6.447272e+00) (55.9877,1.106244e+01) (60.7798,1.218695e+01) (54.8481,1.194351e+01) (70.4354,1.386858e+01) (81.7929,1.802221e+01) (80,2.101980e+01) (99.7138,2.300138e+01) (106.548,2.340058e+01) (107.487,2.214573e+01) (102.871,2.703217e+01) (123.111,2.723235e+01) (105.653,2.599397e+01) (142.617,2.985330e+01) (140.899,2.837679e+01) (120.443,2.692597e+01) (148.974,2.955985e+01) (164.109,3.389623e+01) (147.361,3.247479e+01) (185.915,3.363148e+01) (180,3.486578e+01) (184.424,3.454420e+01) (207.005,3.414667e+01) (227.831,3.470223e+01) (178.041,3.254511e+01) (211.69,3.471495e+01) (234.598,3.587715e+01) (214.882,3.260073e+01) (174.494,3.355162e+01) (252.895,3.885142e+01) (240.665,3.723480e+01) (242.131,3.502036e+01) (299.144,3.699485e+01) (253.546,3.645007e+01) (252.771,3.522147e+01) (246.084,3.627945e+01) (271.997,3.669687e+01) (302.643,3.247440e+01) (300,3.757411e+01) (358.056,3.693197e+01) (275.085,3.448800e+01) (293.539,3.329187e+01) (326.368,3.397828e+01) (322.093,3.598461e+01) (331.071,3.930501e+01) (336.633,3.424821e+01) (367.829,4.047495e+01) (312.583,3.464729e+01) (365.903,3.855801e+01) (383.718,3.559640e+01) (402.819,3.679601e+01) (394.828,3.671301e+01) (401.279,3.912479e+01) (381.05,3.915043e+01) (431.539,3.934377e+01) (421.329,4.005324e+01) (395.937,3.739075e+01) (367.438,3.935526e+01) (472.824,3.930308e+01) (461.64,3.748645e+01) (446.297,3.653252e+01) (462.651,3.991757e+01) (469.15,3.893370e+01) (449.853,3.820327e+01) (472.895,4.004406e+01) (522.747,4.110477e+01) (480,4.175215e+01) (517.513,3.946323e+01) (487.234,4.057889e+01) (438.693,3.934736e+01) (536.651,3.970882e+01) (443.631,3.592164e+01) (520,4.056069e+01) (573.836,4.084618e+01) (546.911,4.020834e+01) (557.431,3.967994e+01) (612.133,4.027972e+01) (591.525,4.118905e+01) (480.983,3.875382e+01) (571.887,3.893361e+01) (594.245,3.995270e+01) (463.568,3.970883e+01) (551.103,3.706661e+01) (527.044,3.807442e+01) (564.366,3.607765e+01) (622.545,3.980778e+01) (660.944,3.750647e+01) (559.443,3.684391e+01) (605.135,4.145768e+01) (628.978,4.011588e+01) (676.72,4.063205e+01) (667.23,4.049968e+01) (595.175,4.160023e+01) (663.976,3.927029e+01) (689.254,4.031096e+01) (674.656,4.217777e+01) (721.497,3.715488e+01) (712.527,4.051868e+01) (648.497,3.948662e+01) (598.719,4.155202e+01) (648.769,4.165227e+01) (625.723,4.110746e+01) (680.046,4.095658e+01) (709.969,4.091812e+01) (719.846,3.967427e+01) (742.657,3.788810e+01) (780.952,4.074511e+01) (755.541,4.120229e+01) (823.626,4.041140e+01) (759.464,3.947798e+01) (783.98,4.023931e+01) (775.648,3.821203e+01) (809.406,4.199662e+01) (752.043,4.182247e+01) (886.345,4.179042e+01) (820,4.064121e+01) (841.118,3.876770e+01) (921.067,3.937417e+01) (907.155,3.939046e+01) (872.71,4.238031e+01) (812.542,4.128160e+01) (779.197,4.006444e+01) (810.983,4.145093e+01) (865.464,4.180096e+01) (760.191,3.896063e+01) (801.067,4.075010e+01) (914.491,4.100355e+01) (891.207,4.012791e+01) (775.183,4.113861e+01) (902.99,4.067157e+01) (862.313,4.172376e+01) (870.549,4.046752e+01) (926.322,4.245962e+01) (934.636,4.072353e+01) (935.479,4.033626e+01) (933.285,4.104386e+01) (862.758,4.003340e+01) (862.169,4.278145e+01) (990.766,4.209830e+01) (980,4.226666e+01) (966.931,3.853926e+01) (973.376,4.182508e+01) (1009.3,4.081928e+01) (1002.33,4.150171e+01) (1000,4.300549e+01) (1013.29,4.190831e+01) (1020,4.185320e+01) (938.189,4.135495e+01) (901.229,4.064124e+01) (909.775,4.089510e+01) (983.281,3.643521e+01) (1049.9,4.138278e+01) (956.978,4.158938e+01) (997.283,4.123964e+01) (1140.58,4.141745e+01) (1061.69,4.205554e+01) (1004.78,4.244843e+01) (1124.21,4.163690e+01) (1056.68,4.038411e+01) (1111.01,4.008347e+01) (1077.48,4.066748e+01) (1317.61,3.892729e+01) (1084.75,4.009116e+01) (1223.27,3.977547e+01) (1252.52,4.192777e+01) (1159.66,4.107418e+01) (1110.91,3.968059e+01) (1115.18,4.067159e+01) (1057.18,4.246233e+01) (1163.78,4.105183e+01) (1130.13,4.115480e+01) (1274.84,4.016729e+01) (1253.94,4.134136e+01) (1299.29,4.108024e+01) (1249.9,4.114512e+01) (1209.24,4.204800e+01) (1166.66,3.805706e+01) (1180.39,4.170787e+01) (1226.52,4.215753e+01) (1234.15,3.974066e+01) (1240,4.212198e+01) (1241.26,3.871423e+01) (1266.96,4.166779e+01) (1244.95,3.972547e+01) (1306.67,3.933705e+01) (1344.31,4.105425e+01) (1230.15,4.174126e+01) (1251.97,3.969564e+01) (1333.78,3.982723e+01) (1175.54,4.161395e+01) (1311.95,4.077019e+01) (1304.49,4.160540e+01) (1376.24,4.210087e+01) (1402.38,4.139451e+01) (1346.98,4.071784e+01) (1289.14,3.831103e+01) (1270.97,4.105580e+01) (1411.18,4.053074e+01) (1326.39,4.044934e+01) (1370.94,4.080128e+01) (1350.02,3.928975e+01) (1391.76,4.198604e+01) (1307.14,4.138135e+01) (1413.3,3.990142e+01) (1368.29,4.049539e+01) (1438.43,4.213186e+01) (1420,4.163076e+01) (1404.61,3.942241e+01) (1454.52,4.137485e+01) (1384.6,3.898644e+01) (1405.34,4.131131e+01) (1499.2,4.109575e+01) (1336.38,3.899214e+01) (1454.93,3.953327e+01) (1410.49,4.268762e+01) (1495.53,3.873049e+01) (1509.18,4.105500e+01) (1500,4.206020e+01) (1552.79,4.246979e+01) (1508.62,4.164846e+01) (1440.28,4.102996e+01) (1411.34,4.159936e+01) (1579.88,4.258924e+01) (1504.97,3.867109e+01) (1597.8,4.157013e+01) (1519.33,4.083860e+01) (1595.11,4.076095e+01) (1323.69,3.949145e+01) (1543.46,4.207946e+01) (1589.94,4.129485e+01) (1532.66,4.134533e+01) (1429.7,4.118308e+01) (1603.33,4.089454e+01) (1652.9,4.176974e+01) (1610.6,4.003639e+01) (1502.79,4.114766e+01) (1506.71,4.104551e+01) (1727.59,4.168705e+01) (1430.36,4.075723e+01) (1450.79,4.154245e+01) (1605.08,4.000385e+01) (1532.06,4.137353e+01) (1626.39,4.100842e+01) (1442.87,4.014673e+01) (1785.19,3.983995e+01) (1635.15,4.167629e+01) (1724.96,3.973470e+01) (1563.28,3.910301e+01) (1841.27,4.177554e+01) (1484.85,4.235303e+01) (1684.06,4.071097e+01) (1727.97,3.704856e+01) (1356.41,4.148328e+01) (1369.9,3.736245e+01) (1769.2,4.119403e+01) (1740.12,4.048728e+01) (1760,4.230415e+01) (1616.64,4.036296e+01) (1578.33,4.187619e+01) (1671.52,4.195361e+01) (1734.06,4.088301e+01) (1743.25,4.106248e+01) (1875.77,4.192812e+01) (1717.29,4.077426e+01) (1735.03,4.158366e+01) (1806.64,3.969305e+01) (1817.32,3.945182e+01) (1784.87,4.091891e+01) (1967.47,4.059420e+01) (1710.68,3.910214e+01) (2003.11,4.129873e+01) (1919.73,4.118041e+01) (1857.18,4.145089e+01) (1814.81,4.114176e+01) (1802.15,4.170184e+01) (1909.88,3.676190e+01) (1934.77,3.494258e+01) (1642.93,4.002935e+01) (1771.45,4.173543e+01) (1900,4.185839e+01) (1885.43,3.949845e+01) (1965.92,3.890764e+01) (1970.12,4.168331e+01) (1937.08,3.948041e+01) (1969.24,4.189589e+01) (1833.98,4.115672e+01) (1800.81,3.994732e+01) (2229.07,4.082654e+01) (1870.66,4.187333e+01) (1915.21,4.052685e+01) (1744.02,4.085663e+01) (1987.9,4.084587e+01) (1902.77,4.076840e+01) (2049.37,4.197720e+01) (1940.94,4.156586e+01)};
\addplot [mark=star,draw=none,color=violet] coordinates {(18.315428,0.008149) (21.253171,0.021868) (18.566355,0.015478) (30.779567,0.071558) (38.619575,0.142574) (40.655854,0.171867) (55.987667,0.433869) (60.779821,0.513216) (54.848066,0.407407) (70.435382,0.868174) (81.792854,1.200000) (80.000000,1.316195) (99.713849,1.963248) (106.548480,2.623861) (107.487282,2.576465) (102.871427,2.494021) (123.111231,3.352956) (105.653072,2.606320) (142.617427,4.564595) (140.898977,4.604444) (120.442808,3.422527) (148.974336,5.780140) (164.108594,6.155588) (147.361260,5.451448) (185.914633,9.562500) (180.000000,8.750188) (184.423724,7.612427) (207.005111,10.166648) (227.830900,10.535412) (178.040979,6.762910) (211.690066,12.802159) (234.597833,11.861608) (214.881853,11.313569) (174.493775,7.194313) (252.894778,16.605832) (240.664823,14.844728) (242.131022,13.107590) (299.144228,15.424719) (253.546051,13.171200) (252.771299,18.574353) (246.084445,12.003441) (271.997261,15.221664) (302.643308,15.455813) (300.000000,21.011673) (358.056190,20.621887) (275.085373,12.148381) (293.538883,16.197759) (326.367884,20.088645) (322.093051,22.746903) (331.070901,16.345946) (336.633442,18.824574) (367.828528,24.765564) (312.583406,17.363286) (365.902688,26.009450) (383.717892,24.955124) (402.819151,20.829430) (394.827570,21.490642) (401.279238,21.396026) (381.049729,21.259558) (431.538602,20.352924) (421.329123,20.413100) (395.936527,19.460505) (367.438100,10.632965) (472.823803,20.008631) (461.640019,18.816257) (446.297321,18.805602) (462.651355,27.634659) (469.150116,22.244894) (449.853187,19.454181) (472.895361,22.320304) (522.746806,27.203961) (480.000000,28.561983) (517.512777,15.358192) (487.234175,18.807805) (438.693092,22.208918) (536.650876,24.684926) (443.630567,19.834184) (520.000000,24.549629) (573.835627,24.642254) (546.911170,32.412839) (557.431262,26.765109) (612.132997,22.555846) (591.525307,25.946596) (480.983401,22.504423) (571.887046,25.154784) (594.244976,22.366660) (463.567845,18.737606) (551.103460,22.653908) (527.044184,23.857248) (564.365539,27.239764) (622.544933,23.049905) (660.944103,36.080165) (559.442677,22.715645) (605.135184,33.157803) (628.977935,23.735584) (676.719610,20.070811) (667.229609,27.687703) (595.174630,25.762864) (663.976000,27.371378) (689.253888,29.861336) (674.656154,36.897221) (721.496886,22.181060) (712.526869,26.002444) (648.497104,24.932530) (598.719491,34.292562) (648.768635,16.251098) (625.723416,25.959137) (680.046133,22.392823) (709.968920,32.894935) (719.845893,28.728311) (742.657114,24.810249) (780.952210,23.639671) (755.540537,24.508120) (823.625745,25.303514) (759.464073,25.447933) (783.979661,26.683619) (775.647553,25.796852) (809.405650,35.251587) (752.042551,37.676606) (886.345321,33.039311) (820.000000,35.296588) (841.118126,32.088975) (921.067444,26.658079) (907.154822,27.825294) (872.709744,29.467125) (812.542076,24.331472) (779.197070,35.090343) (810.983351,22.851586) (865.463665,25.797125) (760.190719,25.945344) (801.066685,35.092383) (914.491341,37.939478) (891.206544,33.818590) (775.183135,25.080194) (902.990055,35.719691) (862.313583,28.809705) (870.549000,38.033632) (926.321679,36.176436) (934.636117,40.032675) (935.479000,35.278588) (933.285485,38.750691) (862.757473,31.093709) (862.168962,34.012506) (990.765982,39.302215) (980.000000,32.619117) (966.931259,35.188223) (973.376422,37.432241) (1009.299910,36.321758) (1002.327910,39.193555) (1000.000000,30.384516) (1013.289279,31.162763) (1020.000000,36.288680) (938.188756,33.467426) (901.228544,30.802451) (909.775141,28.875947) (983.280855,27.666909) (1049.904148,38.461549) (956.978346,31.486958) (997.283293,25.593064) (1140.582528,33.275766) (1061.691378,27.736671) (1004.775980,32.448746) (1124.214563,27.794366) (1056.681636,34.817642) (1111.010632,34.947376) (1077.483854,25.594958) (1317.608157,33.341661) (1084.745884,37.928590) (1223.265996,37.475850) (1252.522056,29.824028) (1159.659036,30.416602) (1110.910030,38.128277) (1115.175249,37.621213) (1057.180498,35.607241) (1163.778011,33.630554) (1130.132326,22.955263) (1274.843841,35.030144) (1253.939147,36.284884) (1299.291132,34.211595) (1249.896525,31.209034) (1209.238690,37.513534) (1166.661709,29.519230) (1180.394869,28.680228) (1226.520672,34.338036) (1234.153431,26.573487) (1240.000000,34.751189) (1241.255593,32.912637) (1266.961467,32.977924) (1244.949512,31.088120) (1306.673627,34.702545) (1344.306690,36.072548) (1230.145972,31.983121) (1251.967121,39.233476) (1333.781238,39.656377) (1175.541926,39.457858) (1311.951019,39.070712) (1304.485022,38.291533) (1376.236754,39.736431) (1402.383965,38.659167) (1346.980489,40.369397) (1289.135386,38.756736) (1270.965022,32.519538) (1411.175185,38.750919) (1326.394592,32.528189) (1370.940658,42.264120) (1350.020082,36.844084) (1391.760363,33.069013) (1307.138637,39.437971) (1413.298750,37.370406) (1368.287932,36.791327) (1438.426522,38.072781) (1420.000000,43.703978) (1404.606247,40.345249) (1454.519769,40.256967) (1384.599649,30.771103) (1405.338421,38.927479) (1499.199573,40.769510) (1336.377772,39.569011) (1454.925304,38.228904) (1410.491092,39.397570) (1495.530160,36.977994) (1509.182078,39.618942) (1500.000000,44.118224) (1552.787215,37.520293) (1508.617069,41.570565) (1440.277724,40.256166) (1411.335843,38.587501) (1579.877788,38.538714) (1504.966648,35.003810) (1597.803235,40.010711) (1519.333422,39.184946) (1595.107534,39.664593) (1323.691960,38.709537) (1543.455937,33.811266) (1589.937368,42.121853) (1532.658099,39.498587) (1429.700167,42.050881) (1603.326413,38.740119) (1652.898487,38.999249) (1610.596335,39.519850) (1502.789051,37.784507) (1506.705975,40.259272) (1727.590181,40.583474) (1430.355717,35.992519) (1450.793889,37.801638) (1605.083002,38.118764) (1532.059255,44.059619) (1626.392463,39.382953) (1442.867550,39.951295) (1785.189698,39.325263) (1635.146467,36.624300) (1724.957610,39.280410) (1563.275131,38.133417) (1841.268816,40.422460) (1484.853799,37.296483) (1684.058758,43.586333) (1727.972222,38.909320) (1356.408053,36.425409) (1369.897682,35.214434) (1769.199216,39.425110) (1740.117059,38.879510) (1760.000000,44.556853) (1616.639684,42.018899) (1578.328200,37.440946) (1671.523536,42.894671) (1734.057360,42.275617) (1743.246329,41.492577) (1875.770180,38.829245) (1717.291121,39.550365) (1735.034595,42.041718) (1806.642127,39.980880) (1817.317477,40.396229) (1784.873234,39.940295) (1967.469072,39.935628) (1710.677233,40.944276) (2003.113019,40.290922) (1919.728152,39.983735) (1857.176862,31.599119) (1814.814406,40.093267) (1802.154457,42.003268) (1909.881403,39.013777) (1934.771071,42.189632) (1642.925613,38.690529) (1771.449545,39.648002) (1900.000000,38.546485) (1885.433932,39.026078) (1965.920940,40.053771) (1970.121844,37.699753) (1937.079454,41.720000) (1969.244899,40.406708) (1833.978068,38.602837) (1800.804973,39.228360) (2229.069931,36.484041) (1870.655913,43.923590) (1915.212290,41.356185) (1744.022679,33.470225) (1987.895745,41.827184) (1902.770658,41.274411) (2049.371213,37.837838) (1940.939779,40.638703)};
\legend{BLIS, BSMTC, TTT}
\end{axis}
\end{tikzpicture}

}\subfloat[Multi-core performance (twelve cores).]{\thispagestyle{empty}
\begin{tikzpicture}
\pgfplotsset {grid style={dotted,color=black}}
\begin{axis}[
scale=0.8,
grid=major,
ylabel shift=-5,
ymin=0,
ymax=56,
ytick distance=10,
yticklabels=,
xmin=0,
xmax=5000,
scaled x ticks=real:1000,
xlabel={$\tilde m=\tilde n=\tilde k$},
legend style={legend pos=north east,font=\footnotesize},
legend columns=3
]
\addplot [mark=o,color=blue] coordinates {(50,1.20445) (100,6.55625) (150,13.437) (200,20.9252) (250,25.621) (300,26.2455) (350,29.4137) (400,31.0896) (450,30.5823) (500,32.3556) (550,31.6517) (600,33.2574) (650,33.4505) (700,34.1791) (750,35.7891) (800,34.9779) (850,35.6075) (900,36.2398) (950,35.9722) (1000,36.5769) (1050,36.1169) (1100,36.3017) (1150,36.9929) (1200,36.6677) (1250,36.7786) (1300,35.4206) (1350,35.9107) (1400,36.6179) (1450,36.3118) (1500,37.2611) (1550,36.671) (1600,36.8365) (1650,37.0576) (1700,36.5603) (1750,36.7045) (1800,36.5539) (1850,36.2163) (1900,36.8342) (1950,36.7075) (2000,36.6828) (2050,34.3866) (2100,36.109) (2150,36.7212) (2200,36.8887) (2250,37.1752) (2300,37.2444) (2350,36.6638) (2400,37.011) (2450,37.4444) (2500,37.5473) (2550,37.823) (2600,36.6579) (2650,36.9613) (2700,37.2547) (2750,37.4037) (2800,37.2989) (2850,37.1622) (2900,37.4888) (2950,38.1) (3000,37.882) (3050,38.1529) (3100,37.1992) (3150,37.8358) (3200,37.7505) (3250,38.1366) (3300,38.111) (3350,37.0256) (3400,36.2595) (3450,37.5812) (3500,37.9766) (3550,37.2306) (3600,37.2928) (3650,37.4127) (3700,38.0933) (3750,38.4178) (3800,38.4586) (3850,37.9269) (3900,37.5117) (3950,37.653) (4000,36.3821) (4050,37.5811) (4100,34.1292) (4150,36.5366) (4200,36.6256) (4250,37.309) (4300,37.1202) (4350,37.3059) (4400,36.585) (4450,37.3145) (4500,37.3261) (4550,37.0979) (4600,37.1519) (4650,37.252) (4700,37.1362) (4750,36.2753) (4800,36.5848) (4850,36.2192) (4900,36.8889) (4950,37.3259) (5000,37.5297)};
\addplot [mark=triangle,draw=none,color=brown!80!black] coordinates {(51.2993,0.612262) (49.3242,0.48815) (50.606,0.512055) (89.1958,2.43228) (91.6496,2.43706) (99.3288,3.33204) (149.277,8.37399) (145.634,7.55491) (158.687,8.91985) (195.806,13.6549) (217.549,14.5244) (189.958,13.1624) (243.288,17.6539) (247.246,18.6124) (265.863,20.4155) (296.674,20.1444) (275.84,18.4004) (313.043,19.7855) (332.49,21.2189) (371.693,24.853) (335.805,22.7858) (400,27.9987) (387.734,24.4008) (386.838,25.7037) (443.232,28.2179) (479.233,26.5257) (362.428,24.4958) (505.309,30.5325) (360.236,24.2047) (516.5,29.1657) (444.252,27.4079) (536.865,28.7631) (592.613,30.7985) (583.553,28.8609) (579.294,31.4009) (556.991,31.9593) (588.158,29.4082) (642.886,31.8661) (621.429,31.0378) (641.992,30.6179) (735.478,33.1246) (671.524,32.1506) (682.451,32.4754) (744.641,33.1357) (749.333,34.033) (826.449,31.1972) (795.705,32.38) (792.259,33.3297) (930.701,32.3475) (876.159,33.8806) (821.737,33.2931) (900,34.5239) (854.756,33.4449) (851.303,32.4778) (953.99,33.328) (949.546,33.8353) (948.665,33.4917) (1038.5,35.5652) (1010.38,31.8818) (1002.96,34.5561) (1007.57,33.3854) (1081.44,33.0553) (1050,34.716) (1106.63,35.7032) (1119.81,30.9974) (1172.33,34.1588) (1234.17,34.9312) (1210.01,35.0994) (1134.8,35.4493) (1192.66,33.0478) (1256.01,35.4514) (1200,35.5677) (1255.9,35.055) (1053.13,33.1786) (1166.71,34.3932) (1293.09,34.5789) (1312.55,33.7082) (1299.35,34.9719) (1248.36,34.1579) (1402.45,32.5012) (1404.74,32.7705) (1365.35,34.2861) (1256.21,34.6681) (1300.06,33.0573) (1448.99,35.4042) (1441.07,34.0652) (1431.49,34.6651) (1572.44,35.0986) (1597.05,34.5747) (1615.48,33.7283) (1648.41,32.8436) (1542.63,35.6473) (1370.31,34.7303) (1578.38,34.4974) (1601.71,35.3858) (1599.67,32.6113) (1741.5,34.9926) (1572.41,34.0811) (1383.12,35.117) (1615.54,36.0247) (1759.03,33.2105) (1805.3,33.0787) (1636.53,30.2256) (1772.38,35.3128) (1752.59,35.3474) (1834.78,34.0383) (1748.54,34.7085) (1607.5,34.1885) (1833.18,34.4799) (1853.99,35.2039) (1868.56,34.8518) (1933.77,35.6021) (1902.39,36.1379) (1888.01,32.9783) (1979.68,35.3942) (1912.82,35.4267) (1998.77,35.115) (2172.34,35.1195) (1856.92,33.5042) (2142.56,36.1233) (1812.49,29.7829) (1651.84,34.9421) (1986.94,31.8326) (2109.08,33.9181) (1631.94,32.9311) (1979.56,34.8126) (2009.08,34.9716) (1997.42,36.4177) (1982.8,35.7415) (1940.25,35.8645) (2202.49,34.994) (2091.39,34.5255) (2263.97,35.6068) (2175.31,34.8435) (2084.67,36.3786) (2542.69,31.2629) (2300,36.2083) (2068,36.2855) (2353.27,35.5337) (2227.34,34.3153) (2250.94,34.4638) (2361.9,32.1545) (2428.25,32.894) (2374.73,35.2775) (2221.93,36.0914) (2354.28,36.3119) (2389.04,32.7941) (2500,36.7015) (2606.62,35.2058) (2648.9,33.8544) (2273.38,34.7518) (2540.93,36.7757) (2455.51,35.5666) (2600,36.6015) (2470.04,34.4285) (2468.51,31.2872) (2713.15,35.4429) (2547.33,35.1046) (2565.77,35.1288) (2150.65,35.6446) (2806.42,36.6672) (2785.53,35.5533) (2807.38,34.3785) (2628.44,33.1409) (2754.14,35.7801) (2874.66,36.5067) (2699.79,35.9013) (2857.02,34.9912) (2734.01,36.0009) (2813.54,36.7415) (2725.28,36.0695) (2579.14,34.7805) (2887.28,35.6046) (2857.48,36.1746) (2970.38,31.9092) (2829.79,35.3498) (3000.3,36.7982) (3036.52,32.8111) (2877.03,37.2296) (2998.66,36.7498) (2750.28,33.2856) (2751.66,36.0248) (2365.94,34.5935) (3125.55,35.8527) (2949.35,37.6241) (3130.37,37.0256) (3135.94,36.4338) (2858.31,35.2869) (2861.48,37.1299) (3266.03,37.8946) (3197.66,36.3198) (2864.12,34.624) (3422.88,34.889) (3004.94,35.2825) (3492.43,35.0884) (3246.47,35.1861) (2574.94,32.9272) (2980.14,36.6962) (3598.63,35.3956) (3028.87,35.8911) (3225.42,35.3691) (3561.06,35.9385) (3117,35.9694) (3673.46,35.0744) (3100.6,35.5296) (2828.13,35.4956) (3664.3,36.4149) (3387.04,34.8876) (3506.76,35.8538) (3459.26,36.6491) (3288.58,35.3016) (3294.18,35.7197) (3527.6,36.7684) (3611.72,37.3408) (3899.69,36.1908) (3171.5,33.87) (3699.5,36.1216) (3620.74,36.4823) (3269.5,33.0953) (3461.48,36.2727) (3644.92,34.3432) (3192.64,34.4726) (3812.3,36.7623) (3739.3,37.1246) (3522.25,35.4267) (3480.76,36.81) (3793.32,37.2795) (3406.58,36.5104) (4245.37,31.6203) (3867.92,36.2235) (3850,36.4834) (3874.58,37.1015) (3855.49,36.7456) (3922.54,35.932) (3929.22,34.0152) (4020.21,37.0829) (3862.05,36.4448) (3915.65,37.0682) (4049.84,33.3069) (3802.84,36.5909) (3915.99,29.6694) (4030.02,35.4468) (3887.57,35.0746) (3917.01,31.6052) (3775.79,30.8059) (4258.97,35.4983) (3877.36,34.4698) (3978.16,33.7675) (4108.47,36.2048) (4602.2,32.8534) (3908,33.7212) (4200,36.9626) (4247.33,36.8337) (3612.8,32.7858) (4137.11,34.0933) (4281.92,36.6451) (4193.68,37.3862) (4157.76,34.7647) (4466.99,34.9029) (4350.59,36.3435) (4307.55,36.8639) (4454.18,36.7406) (3994.53,32.0273) (4119.65,34.1744) (4466.84,35.2334) (4450,36.8007) (4282.66,36.5794) (4500,37.1905) (4481.38,34.0443) (4480.27,34.9191) (4314.71,37.3127) (4299.96,37.3182) (4904.17,36.6522) (4581.37,35.1353) (4396.17,35.2931) (4494.01,37.2436) (4621.49,36.991) (4553.6,35.2017) (4881.61,36.4713) (4353.08,33.912) (4318.06,34.4457) (4713.32,32.7948) (4517.21,33.7676) (4899.89,35.8011) (4521.25,33.8881) (4800,37.1712) (4988.04,33.223) (4885.23,32.5829) (4600.26,35.2462) (4923.51,35.7502) (4831.93,36.6136) (4722.69,34.538) (4745.16,36.48) (5288.97,36.9555) (5180.59,37.0303) (4439.06,33.2874) (4966.13,35.0637) (5333.56,36.8576) (5000,37.6602) (4995.47,36.6242)};
\addplot [mark=star,draw=none,color=violet] coordinates {(51.299278,0.0108277) (49.324241,0.0157604) (50.605960,0.0174053) (89.195798,0.103838) (91.649581,0.11891) (99.328839,0.179487) (149.276516,0.410971) (145.634163,0.372234) (158.687174,0.477077) (195.806000,0.982875) (217.548542,1.30197) (189.957516,0.871396) (243.288080,1.92926) (247.245714,1.64107) (265.862857,2.11479) (296.674382,2.19798) (275.839857,1.51363) (313.042535,2.17751) (332.490252,1.75886) (371.692514,2.05389) (335.805443,2.1952) (400.000000,4.73023) (387.734399,4.09924) (386.838344,3.49819) (443.232055,7.40056) (479.233151,3.33766) (362.428032,2.75023) (505.308768,5.00209) (360.236470,4.16426) (516.500166,5.61483) (444.252109,5.31186) (536.864546,4.1125) (592.613134,8.55616) (583.553296,3.62443) (579.293631,6.04816) (556.990660,5.27376) (588.158098,3.80672) (642.885667,5.0966) (621.429238,6.04088) (641.992488,4.0875) (735.478147,12.943) (671.523968,9.48328) (682.450787,6.52068) (744.640909,11.7795) (749.332740,9.41528) (826.449243,9.22263) (795.705319,12.6265) (792.258665,9.39044) (930.700825,7.12347) (876.159225,5.48747) (821.737344,6.50123) (900.000000,5.38803) (854.756266,11.8693) (851.302593,5.4776) (953.990234,6.03814) (949.546450,15.2042) (948.664791,9.15605) (1038.498820,7.92943) (1010.382497,8.52569) (1002.957909,10.9117) (1007.565656,8.23487) (1081.444651,9.1025) (1050.000000,13.391) (1106.626666,9.87309) (1119.813064,19.8572) (1172.334880,9.39334) (1234.174828,10.0767) (1210.012230,16.1623) (1134.799980,18.0736) (1192.655136,17.39) (1256.011576,20.7047) (1200.000000,12.8807) (1255.897465,9.98609) (1053.133497,16.1807) (1166.706348,16.2604) (1293.086374,10.5124) (1312.545214,8.53632) (1299.353683,18.7325) (1248.360103,10.846) (1402.454029,16.7484) (1404.737285,11.1439) (1365.349461,11.109) (1256.214387,12.5275) (1300.060936,13.6965) (1448.986878,15.2475) (1441.066918,20.491) (1431.488824,8.79523) (1572.444837,13.8595) (1597.051863,21.3457) (1615.477303,9.694) (1648.408677,21.033) (1542.631695,7.51753) (1370.306199,13.3964) (1578.375728,14.2687) (1601.714828,12.9972) (1599.666597,16.7409) (1741.498907,10.5309) (1572.408436,24.6081) (1383.116408,19.8499) (1615.539986,14.2908) (1759.030113,20.5209) (1805.297726,21.636) (1636.531311,19.243) (1772.379257,10.6361) (1752.594248,24.1289) (1834.783508,25.6771) (1748.543005,18.0635) (1607.496334,12.5595) (1833.180887,16.512) (1853.991382,15.9822) (1868.562671,12.0843) (1933.770788,22.76) (1902.385775,11.9361) (1888.006555,16.3595) (1979.682635,20.7159) (1912.816910,10.4898) (1998.770077,9.56308) (2172.336812,12.146) (1856.918194,10.4364) (2142.560187,13.3924) (1812.494898,20.95) (1651.843229,19.1486) (1986.936947,9.92046) (2109.079733,13.5136) (1631.939829,20.625) (1979.557599,17.4761) (2009.081113,17.54) (1997.422012,17.6175) (1982.804108,8.21413) (1940.251145,14.102) (2202.491118,16.4744) (2091.393346,13.8971) (2263.966460,15.9076) (2175.307923,13.0384) (2084.672067,12.8757) (2542.692992,17.2108) (2300.000000,15.1777) (2068.003801,17.3375) (2353.269775,15.4826) (2227.337139,13.8451) (2250.941236,11.5455) (2361.896965,16.5339) (2428.249970,18.0844) (2374.734966,17.1234) (2221.926175,14.3974) (2354.278970,17.0317) (2389.040586,16.846) (2500.000000,13.7314) (2606.619257,17.7088) (2648.903308,18.1826) (2273.384587,16.2897) (2540.927762,18.0165) (2455.506371,15.9379) (2600.000000,14.713) (2470.043326,17.3592) (2468.510677,14.5233) (2713.149844,20.0369) (2547.332112,12.1083) (2565.772391,17.903) (2150.651948,14.9792) (2806.417254,15.5084) (2785.528738,18.7316) (2807.384480,18.9282) (2628.439030,19.2083) (2754.135112,18.451) (2874.658308,16.1394) (2699.786897,17.7475) (2857.017080,18.7027) (2734.010127,19.0901) (2813.535441,14.1793) (2725.283770,16.7387) (2579.142796,15.0168) (2887.277601,21.1682) (2857.483748,16.5784) (2970.376484,19.7378) (2829.792304,18.5883) (3000.304325,18.3831) (3036.524190,18.9419) (2877.028184,16.5053) (2998.655397,18.7819) (2750.279993,19.7368) (2751.659604,20.0776) (2365.942648,13.8517) (3125.552244,21.4833) (2949.353523,14.3258) (3130.368202,18.3183) (3135.937313,15.7525) (2858.310003,18.7949) (2861.480317,18.4241) (3266.032615,20.4692) (3197.664963,21.7229) (2864.122208,18.8123) (3422.882560,20.0265) (3004.938533,19.3258) (3492.428804,19.6825) (3246.467192,20.2263) (2574.942688,18.2953) (2980.135315,18.9166) (3598.632583,18.8735) (3028.866172,13.8932) (3225.424898,20.6767) (3561.058054,21.4184) (3116.997111,17.8635) (3673.455998,20.7724) (3100.598164,19.5769) (2828.131316,18.8074) (3664.304530,22.9726) (3387.035937,20.4331) (3506.764197,21.1418) (3459.260982,20.6878) (3288.584191,19.3675) (3294.178626,21.3323) (3527.603533,20.9959) (3611.721792,22.742) (3899.692283,23.2346) (3171.501337,20.4779) (3699.495468,22.0672) (3620.744401,22.6791) (3269.500808,20.1779) (3461.479252,21.9148) (3644.915520,21.322) (3192.643779,20.512) (3812.302260,22.1888) (3739.302848,19.5218) (3522.254388,18.6343) (3480.755823,20.793) (3793.321603,22.5374) (3406.584379,18.8461) (4245.367964,22.9751) (3867.916494,23.1387) (3850.000000,17.7779) (3874.575470,21.3294) (3855.494040,20.9277) (3922.536190,20.2606) (3929.224251,18.625) (4020.213548,21.147) (3862.051953,21.2631) (3915.648371,22.283) (4049.839998,23.3764) (3802.840351,22.3313) (3915.986432,20.8829) (4030.020633,23.0856) (3887.572870,19.7585) (3917.012934,22.7537) (3775.790325,23.931) (4258.969003,23.3324) (3877.356412,21.4314) (3978.162426,24.3445) (4108.471837,24.4499) (4602.201039,24.1971) (3907.997789,24.2194) (4200.000000,20.2415) (4247.331658,21.8567) (3612.799337,20.9135) (4137.110845,23.9254) (4281.924121,22.948) (4193.679636,23.1533) (4157.762372,24.3787) (4466.988287,22.0281) (4350.586128,25.3106) (4307.546568,23.0583) (4454.183359,21.1125) (3994.531193,23.081) (4119.651249,24.7791) (4466.836593,24.332) (4450.000000,20.67) (4282.656830,21.2074) (4500.000000,20.2075) (4481.378605,25.0771) (4480.269618,26.534) (4314.708957,23.9711) (4299.956531,18.8589) (4904.172927,25.2177) (4581.371036,25.169) (4396.165348,25.3746) (4494.008421,21.4877) (4621.492251,20.8602) (4553.597741,25.315) (4881.606034,21.1836) (4353.081584,23.6469) (4318.057826,23.7201) (4713.323058,24.2796) (4517.206969,23.6331) (4899.887197,22.6804) (4521.249896,25.026) (4800.000000,24.7705) (4988.037536,25.7756) (4885.232786,24.3312) (4600.258656,23.5863) (4923.514443,25.3439) (4831.932780,23.7477) (4722.688691,24.5911) (4745.158502,24.2479) (5288.972205,25.7588) (5180.590388,26.24) (4439.063901,24.3844) (4966.133273,25.5597) (5333.561391,25.8037) (5000.000000,25.0026) (4995.467286,26.5237)};
\legend{BLIS, BSMTC, TTT}
\end{axis}
\end{tikzpicture}

}

\caption{\label{fig:rand-square}Performance of matrix multiplication and randomly
generated tensor contractions for square matrix/tensor shapes on a
Xeon E5-2690 v3 processor.}
\end{figure}

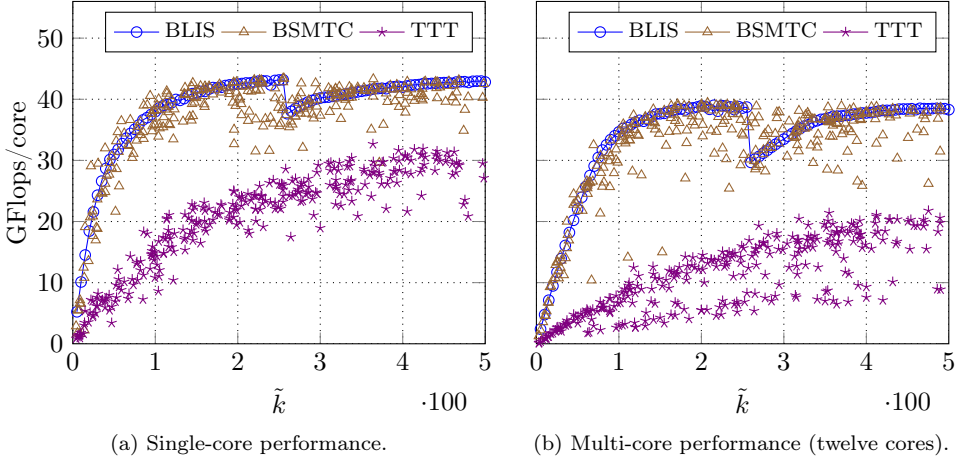
\begin{figure}
\subfloat[Single-core performance.]{\thispagestyle{empty}
\begin{tikzpicture}
\pgfplotsset {grid style={dotted,color=black}}
\begin{axis}[
scale=0.8,
grid=major,
ymin=0,
ymax=56,
ylabel={GFlops/core},
ylabel shift=-5,
ytick distance=10,
xmin=0,
xmax=500,
scaled x ticks=real:100,
xlabel={$\tilde k$},
legend style={legend pos=north east,font=\footnotesize},
legend columns=3
]
\addplot [mark=o,color=blue] coordinates {(5,5.23164) (10,10.1087) (15,14.5177) (20,18.4334) (25,21.5805) (30,24.2995) (35,26.6022) (40,28.6607) (45,30.1288) (50,30.9656) (55,31.7917) (60,32.7378) (65,33.6244) (70,34.4062) (75,35.0815) (80,35.7835) (85,36.6076) (90,36.9858) (95,37.505) (100,37.9833) (105,38.5912) (110,38.9695) (115,39.1446) (120,39.4526) (125,39.9656) (130,39.818) (135,40.1103) (140,40.5849) (145,40.9185) (150,41.0453) (155,41.2833) (160,41.2827) (165,41.6548) (170,41.9081) (175,41.8716) (180,41.9599) (185,42.239) (190,42.3505) (195,42.4054) (200,42.4794) (205,42.569) (210,42.6629) (215,42.6669) (220,42.8288) (225,42.9569) (230,43.0157) (235,42.9743) (240,42.3309) (245,43.0742) (250,43.1505) (255,43.1828) (260,37.6337) (265,38.0239) (270,38.4849) (275,38.8486) (280,39.1989) (285,39.5373) (290,39.8019) (295,40.0513) (300,40.1973) (305,40.3146) (310,40.351) (315,40.4701) (320,40.5599) (325,40.7469) (330,40.8346) (335,40.9413) (340,41.0608) (345,41.1964) (350,41.3045) (355,41.4256) (360,41.5204) (365,41.6897) (370,41.7548) (375,41.7934) (380,41.7294) (385,42.0422) (390,42.0409) (395,42.1091) (400,41.9889) (405,42.1556) (410,42.3104) (415,42.3483) (420,42.4001) (425,42.558) (430,42.5367) (435,42.5795) (440,42.6139) (445,42.6821) (450,42.7161) (455,42.743) (460,42.5915) (465,42.8113) (470,42.8701) (475,42.875) (480,42.7487) (485,42.9341) (490,42.9565) (495,42.9724) (500,42.8192)};
\addplot [mark=triangle,draw=none,color=brown!80!black] coordinates {(7.68115,5.51203) (4,2.8816) (5.57003,1.75368) (8.502,6.61371) (9.78978,6.5735) (6.128,5.46321) (13.875,10.8134) (14.1859,2.20653) (8.625,7.15421) (19,13.5569) (13.1007,12.4818) (8.80209,1.63066) (27.7431,16.9459) (29.835,18.8101) (24.5754,17.661) (28.4925,18.9974) (21,20.6963) (24.3,20.8492) (14.8261,19.108) (24.0275,21.6812) (51.7233,21.6315) (39.2931,26.206) (36,24.5791) (29.7297,23.6398) (44.5725,29.5866) (30.655,29.4912) (21.9898,29.0608) (53.0337,28.6029) (39.1732,31.9951) (34.776,24.6476) (47.088,28.8061) (56.088,25.7775) (44.055,27.6959) (68.16,32.3823) (41.041,27.4983) (67.005,32.5505) (39.2616,24.8404) (82.152,36.3144) (85.8,33.5351) (72.0279,30.4235) (62.37,34.4183) (77.7053,34.896) (47.304,30.6456) (78.4245,35.5182) (55.521,31.4222) (58.8744,36.4442) (54.4907,34.6469) (77.3373,33.2661) (85.0199,34.645) (121.866,33.4678) (82,34.1715) (96,37.6454) (96.525,37.0265) (73.482,31.1045) (93.781,40.5706) (112.292,38.3423) (95,37.6142) (98,38.1261) (117.843,35.7946) (110.298,37.0937) (99.576,35.8374) (52.6048,33.1504) (104.633,35.8713) (103.769,33.9151) (110,39.1098) (99,38.2129) (105.225,35.2704) (96.6,35.5912) (133.65,36.37) (109.538,35.4444) (116.348,35.9714) (145.593,37.2679) (160.272,42.111) (147.042,40.407) (85.9513,38.99) (136,40.7009) (143.942,36.5441) (69.717,36.6312) (90.2409,39.0628) (102.144,39.0081) (135,40.89) (123.232,38.3728) (189.803,40.8738) (115.851,37.9201) (104.098,40.0163) (113.88,41.3289) (122.342,36.9113) (150,41.5002) (149.625,40.2131) (175.395,41.9811) (102,38.2506) (165.927,41.8936) (136.08,41.8944) (116.043,38.8006) (144.196,42.0402) (96.3483,34.793) (197.383,38.9389) (212.565,36.4849) (159.994,40.5369) (144.518,39.2444) (158.086,41.5645) (221.822,36.7339) (213.981,38.0011) (245.203,41.8035) (216.283,41.2754) (169.965,40.8087) (141.703,37.1548) (120,39.8579) (204.324,39.239) (116.821,40.5461) (233.436,36.7871) (177.84,42.4295) (190,41.9381) (189.572,41.3753) (202.549,39.9853) (172.575,41.5686) (180.619,41.1228) (159.3,39.1965) (154.845,42.1656) (198.45,41.1049) (159.762,38.7684) (161.28,41.7379) (278.83,32.0886) (195.515,35.0432) (195,42.6801) (201.792,40.3029) (193.945,39.4793) (195.514,32.682) (165.24,42.1191) (145.47,42.6722) (198.576,42.812) (254.898,42.802) (220,43.1625) (209.588,42.2356) (221.906,41.6011) (238.286,31.5867) (205.632,42.0566) (228.62,43.1131) (230.868,35.5772) (212.355,41.3799) (227.16,41.4359) (113.981,40.9351) (276.463,41.0124) (273.42,38.1957) (245.304,40.1825) (228.284,42.7382) (230.01,42.5906) (256,43.379) (262.414,35.4431) (255,43.41) (221.801,41.4626) (229.27,42.8814) (140.494,40.9119) (220,42.9701) (243.055,36.04) (325.065,36.8025) (135.926,40.7315) (221.469,31.4914) (292.053,36.8843) (270,37.1934) (179.591,42.1554) (312,40.5009) (300.019,37.3317) (296,39.936) (204.538,37.4251) (280,38.4802) (253.884,38.239) (352.95,41.8497) (337.758,41.3653) (264.262,36.3805) (153,41.4311) (286,39.1123) (226.398,43.1493) (410.13,41.6498) (272.791,39.4473) (285,38.8256) (286.15,39.871) (195.053,42.7495) (336.672,41.0636) (422.712,40.8764) (311.6,37.9479) (338.256,40.2943) (147.735,39.4794) (266.023,39.8573) (252.935,39.9568) (239.67,36.9885) (319.095,40.7284) (319.41,40.1648) (282.646,37.6627) (254.826,38.5702) (252.11,36.0269) (396,40.7036) (334.831,40.2351) (273.488,40.1866) (328.784,40.2221) (330.577,39.9425) (379.913,40.9726) (336,40.9695) (317.077,41.0477) (276.226,37.6832) (285.635,39.681) (371.349,41.6656) (381.48,39.5797) (345.84,41.5638) (353.28,41.3627) (297.54,41.3266) (361.887,35.0608) (388.8,41.6729) (333.829,38.8074) (362.6,41.3293) (360,41.7683) (277.014,42.6146) (385,42.2544) (266.22,35.8966) (409.901,40.3606) (405.72,41.0716) (375.12,41.6626) (365,41.8691) (515.99,42.5831) (323.136,42.0456) (270,37.2041) (249.225,38.6324) (394.889,41.7905) (322,40.6225) (374.207,41.7233) (258.048,37.43) (345.42,41.9295) (338.01,41.6338) (366,41.9223) (460.424,40.349) (323.645,41.0295) (601.77,42.4394) (403.17,37.8488) (295.865,38.8311) (536.094,42.828) (310.11,41.8317) (250.404,35.6267) (364.32,43.3007) (290.464,37.5298) (294.354,42.6424) (498.28,42.5296) (525.363,38.2069) (335.34,42.0159) (465.328,41.1371) (453.72,42.2153) (473.868,34.8713) (496.883,40.266) (354.24,41.4875) (395.52,38.771) (424.625,40.5864) (403.2,41.8499) (423.675,41.7451) (421.2,40.8581) (319.008,38.8951) (414.152,39.7627) (456,42.9116) (464.4,39.498) (318.603,41.7159) (424.125,39.8356) (521.64,37.8188) (566.669,41.9778) (440,42.8499) (445.05,41.9564) (405.499,42.8831) (376.512,35.8457) (402.853,39.2989) (440.909,41.843) (318.533,41.7201) (475.38,38.3024) (392.84,42.3639) (389.032,41.6133) (460.084,40.3757) (426,42.7698) (608,42.3513) (371.598,29.6261) (439.04,42.6513) (468.72,42.8382) (451.949,39.6179) (617.694,41.7936) (557.2,39.9381) (532.541,41.5272) (441.597,40.196) (390.74,39.706) (512.198,40.6876) (462.384,42.3242) (442.26,42.7713) (513.708,40.4182) (504,42.894) (427.35,42.903) (501.12,41.5188) (363.842,40.4678) (468,43.1403) (332.801,40.9061) (312.887,41.4099) (571.05,41.4158) (513,39.1338) (510.654,42.6851) (553.609,41.1741) (479.317,40.3464) (422.159,40.1036)};
\addplot [mark=star,draw=none,color=violet] coordinates {(7.681149,1.353261) (4.000000,0.848171) (5.570032,1.125209) (8.502000,0.959776) (9.789780,1.778263) (6.128000,1.383754) (13.875000,2.991591) (14.185884,3.440384) (8.625000,1.387918) (19.000000,3.536920) (13.100724,1.641650) (8.802090,1.465328) (27.743063,5.026315) (29.835000,6.146319) (24.575400,4.659092) (28.492500,6.146918) (21.000000,5.081823) (24.300000,4.423309) (14.826094,5.377009) (24.027500,6.966701) (51.723240,6.903998) (39.293100,6.963196) (36.000000,7.162041) (29.729700,7.562043) (44.572500,5.496416) (30.654956,8.180586) (21.989829,7.174935) (53.033750,7.546871) (39.173148,5.940501) (34.776000,6.960812) (47.088000,3.421052) (56.088000,9.321734) (44.055000,7.956160) (68.160000,8.166236) (41.041000,7.469639) (67.005000,7.366045) (39.261600,7.270835) (82.152000,13.467541) (85.800000,11.227014) (72.027900,10.543497) (62.370000,10.153743) (77.705250,13.359882) (47.304000,6.055542) (78.424500,10.407813) (55.521000,10.548430) (58.874400,8.796866) (54.490725,12.128084) (77.337325,13.084666) (85.019922,7.260144) (121.866120,10.404486) (82.000000,12.032889) (96.000000,15.167899) (96.525000,14.225252) (73.482000,9.645087) (93.780960,12.268571) (112.292180,17.193308) (95.000000,14.609906) (98.000000,13.693785) (117.842850,17.125521) (110.298237,14.837434) (99.576000,10.824361) (52.604842,9.610001) (104.632762,11.044274) (103.768500,8.609194) (110.000000,14.753281) (99.000000,15.046808) (105.225000,12.048434) (96.600000,15.905407) (133.650000,18.648458) (109.538325,17.432173) (116.348400,15.874327) (145.592760,16.914148) (160.272000,21.803860) (147.042000,18.106042) (85.951288,18.317469) (136.000000,17.306102) (143.942400,18.563258) (69.717000,14.277128) (90.240885,18.005751) (102.144000,12.915622) (135.000000,17.036649) (123.232000,18.052830) (189.803376,20.520116) (115.851450,18.581292) (104.098000,16.758999) (113.880000,20.372093) (122.341752,15.069966) (150.000000,19.959333) (149.625000,19.548762) (175.395000,21.242459) (102.000000,14.319495) (165.926880,19.550926) (136.080000,15.336414) (116.042625,17.802812) (144.196200,20.187682) (96.348285,14.712259) (197.382528,22.386269) (212.564756,18.552708) (159.993750,18.927242) (144.518419,22.801896) (158.086080,22.274469) (221.821950,20.075747) (213.981075,18.017468) (245.203200,22.059947) (216.282938,23.494251) (169.965000,20.278363) (141.702912,17.933460) (120.000000,15.807934) (204.324120,22.578888) (116.821320,21.062288) (233.436375,23.429668) (177.840000,21.160722) (189.999810,21.492811) (189.572500,20.212241) (202.548937,19.880336) (172.574820,21.535931) (180.618750,22.588285) (159.300000,23.068464) (154.845000,16.468821) (198.450000,21.877035) (159.761869,20.103023) (161.280000,18.430618) (278.830188,23.397315) (195.515320,21.952674) (195.000000,20.701186) (201.791832,23.970224) (193.945050,22.839484) (195.514087,24.887033) (165.240000,22.113359) (145.470000,21.919274) (198.576000,19.496728) (254.898000,24.178638) (220.000000,24.265821) (209.587500,23.233128) (221.906250,24.272774) (238.286344,26.305367) (205.632000,23.811627) (228.620000,24.168376) (230.868000,19.144028) (212.355000,24.539604) (227.160000,21.924723) (113.981246,20.552553) (276.462588,23.265448) (273.420000,24.782509) (245.303953,23.060369) (228.283650,24.236426) (230.010000,22.949793) (256.000000,25.869201) (262.414152,24.224289) (255.000000,23.208389) (221.801250,22.082616) (229.270500,24.074106) (140.493750,20.640736) (220.000000,24.295801) (243.054630,26.767486) (325.064644,22.568546) (135.926263,21.143190) (221.469188,21.990852) (292.053300,21.015990) (270.000000,26.996119) (179.591037,22.787758) (312.000000,28.346725) (300.019500,18.940737) (296.000000,24.616088) (204.537850,24.145210) (280.000000,26.387400) (253.883700,26.056809) (352.950000,28.574175) (337.758120,26.736706) (264.261690,17.460375) (153.000000,19.918714) (286.000000,24.773635) (226.397500,25.783206) (410.130000,30.001852) (272.790787,25.231175) (285.000000,24.706074) (286.150000,25.685057) (195.053040,25.042216) (336.672000,26.509280) (422.712000,24.223836) (311.600000,25.990627) (338.256000,23.338613) (147.735000,22.464183) (266.023012,26.684690) (252.934812,26.138152) (239.669719,23.120822) (319.095000,25.846285) (319.410000,25.411196) (282.646250,24.518770) (254.826000,25.223429) (252.110232,22.090105) (396.000000,28.418099) (334.831250,25.138425) (273.487500,24.883354) (328.784137,24.469990) (330.577335,26.018124) (379.912500,22.095684) (336.000000,29.095003) (317.077500,28.732418) (276.225768,25.745189) (285.635489,27.011166) (371.349000,27.977379) (381.480000,26.129270) (345.840000,25.079952) (353.280000,28.167055) (297.540000,29.216508) (361.886616,24.078120) (388.800000,30.343368) (333.828936,26.654871) (362.600000,28.134972) (360.000000,27.634050) (277.013803,26.168016) (385.000000,27.104504) (266.220000,24.236176) (409.901280,27.782795) (405.720000,30.461961) (375.120000,29.662768) (365.000000,27.994353) (515.989980,28.373459) (323.136000,29.556276) (270.000000,24.491462) (249.224850,22.313731) (394.888824,31.653384) (322.000000,28.368872) (374.206500,26.125980) (258.048000,28.590794) (345.420000,20.874847) (338.010000,30.383092) (366.000000,29.109346) (460.424250,27.646414) (323.645438,24.552016) (601.770000,23.210894) (403.170000,28.042768) (295.865089,28.053807) (536.094000,29.535181) (310.110352,29.460143) (250.404000,26.068900) (364.320000,29.748301) (290.464125,23.667524) (294.354000,24.763919) (498.279600,27.113891) (525.363300,28.431731) (335.340000,29.853361) (465.327720,29.696221) (453.720000,31.274319) (473.867550,22.792058) (496.883100,29.493589) (354.240000,26.044426) (395.520000,29.963636) (424.625000,31.882419) (403.200000,30.672749) (423.675000,28.672337) (421.200000,28.080995) (319.008339,26.783861) (414.151920,31.809132) (456.000000,30.589976) (464.400000,27.999992) (318.603000,22.797711) (424.125000,30.992104) (521.640000,31.848949) (566.669250,26.542002) (440.000000,30.990553) (445.050000,29.940336) (405.499500,21.647566) (376.512386,29.762281) (402.853500,30.839131) (440.908875,30.133643) (318.532500,25.776582) (475.380000,23.519655) (392.840000,26.343111) (389.032250,28.212848) (460.083834,29.456971) (426.000000,29.295905) (608.000000,34.186773) (371.598240,27.565357) (439.040000,28.321279) (468.720000,29.177095) (451.948612,29.773617) (617.694000,31.954218) (557.200000,31.163747) (532.541137,31.752684) (441.597312,31.497528) (390.740250,28.193399) (512.198437,30.701071) (462.384000,31.401890) (442.260000,30.639691) (513.708000,31.881463) (504.000000,29.956202) (427.350000,30.552688) (501.120000,32.127711) (363.842160,32.669674) (468.000000,29.360503) (332.800650,29.004605) (312.886800,29.648120) (571.050000,31.313252) (513.000000,33.085101) (510.654375,32.559716) (553.609080,30.444741) (479.316960,20.856105) (422.158620,30.891091)};
\legend{BLIS, BSMTC, TTT}
\end{axis}
\end{tikzpicture}

}\subfloat[Multi-core performance (twelve cores).]{\thispagestyle{empty}
\begin{tikzpicture}
\pgfplotsset {grid style={dotted,color=black}}
\begin{axis}[
scale=0.8,
grid=major,
ylabel shift=-5,
ymin=0,
ymax=56,
ytick distance=10,
yticklabels=,
xmin=0,
xmax=500,
scaled x ticks=real:100,
xlabel={$\tilde k$},
legend style={legend pos=north east,font=\footnotesize},
legend columns=3
]
\addplot [mark=o,color=blue] coordinates {(5,2.39336) (10,4.7695) (15,7.12659) (20,9.46347) (25,11.7002) (30,13.9392) (35,16.0779) (40,18.1978) (45,20.2019) (50,22.0427) (55,23.9523) (60,25.7321) (65,27.5264) (70,29.0625) (75,30.4673) (80,31.6884) (85,32.5402) (90,33.385) (95,34.0224) (100,34.6551) (105,35.2322) (110,35.3754) (115,35.8684) (120,36.239) (125,36.5584) (130,36.4008) (135,36.6739) (140,37.1782) (145,37.5761) (150,37.6275) (155,37.6948) (160,37.9691) (165,38.1624) (170,38.3108) (175,38.4216) (180,38.2424) (185,38.5784) (190,38.5868) (195,38.7291) (200,38.9396) (205,38.7993) (210,38.3518) (215,39.011) (220,38.0503) (225,38.951) (230,38.9211) (235,38.7409) (240,38.3016) (245,38.8052) (250,38.5609) (255,38.7184) (260,29.674) (265,30.2043) (270,30.6619) (275,31.1835) (280,31.6645) (285,32.1551) (290,32.6313) (295,33.0954) (300,33.4751) (305,33.9673) (310,34.3773) (315,34.8165) (320,35.2301) (325,35.6122) (330,35.8282) (335,36.2733) (340,36.4801) (345,36.7067) (350,36.7907) (355,36.9483) (360,37.2272) (365,37.2363) (370,37.2508) (375,37.3715) (380,37.4018) (385,37.5717) (390,37.6308) (395,37.7469) (400,37.7586) (405,37.8185) (410,37.9247) (415,38.0609) (420,38.1724) (425,38.1824) (430,38.181) (435,38.2605) (440,38.3833) (445,38.4002) (450,38.3985) (455,38.4495) (460,38.2972) (465,38.5206) (470,38.4673) (475,38.4444) (480,38.4751) (485,38.5131) (490,38.484) (495,38.4599) (500,38.2923)};
\addplot [mark=triangle,draw=none,color=brown!80!black] coordinates {(3.2835,1.26921) (3,1.28113) (5.02611,2.21992) (11.205,4.58616) (10.4823,4.62428) (7.9655,3.56251) (13.9125,6.79103) (14.8575,6.73906) (12.264,5.34554) (17.802,9.4176) (20.01,10.4497) (16.3294,9.29918) (27.135,10.7196) (22.8222,10.7871) (26.9145,11.9672) (30.3099,13.6282) (24.8395,13.3853) (33.1891,11.2452) (23.2969,12.9089) (29.2565,12.4598) (25.9277,13.3047) (39.44,16.9263) (28.9675,14.4229) (36.8603,15.6647) (47.5282,23.076) (37.0556,20.0034) (44.9888,22.7704) (37.9741,17.8385) (55.8057,21.7821) (50.2781,21.7638) (64,26.4511) (50.5096,23.4497) (59.9625,25.6729) (79.888,25.7118) (62.1375,24.5496) (53.82,25.2553) (68.3304,26.5572) (77.688,30.7717) (61.7636,25.4685) (66.8448,10.3856) (45.1686,24.4817) (35.4578,14.027) (61.1953,26.5816) (79.2321,30.0467) (78.39,27.9725) (105.366,35.9211) (81.2,28.8313) (65.2645,26.3645) (89.0723,25.6574) (81.8642,26.4148) (88.3575,31.2207) (96.6937,31.823) (89.4375,33.2893) (77.4125,21.3355) (116.41,28.2081) (73.3455,29.1395) (61.8739,25.8993) (69.3,32.5179) (100,34.7537) (62.4291,26.2307) (96.642,30.0643) (94.117,27.9395) (139.78,35.6904) (103.764,33.8988) (106.92,35.587) (89.5548,30.662) (107.906,32.8026) (112.664,33.3425) (113.325,33.2766) (98.5411,29.6652) (107.39,33.1573) (118.965,36.3291) (132,37.109) (112.433,35.0568) (122.487,34.2107) (139.104,37.1806) (111.53,34.5064) (127.361,27.5841) (157.245,32.4267) (120,35.9843) (101.812,33.1775) (88.2194,29.886) (141.977,36.2846) (152.558,14.9886) (143.769,36.0967) (168,38.5988) (110.893,14.1277) (166.898,37.4909) (148.803,34.3694) (104.247,34.3041) (132.962,36.4978) (113.636,34.1378) (155,38.1425) (150,37.6327) (167.034,38.2791) (143.872,35.0811) (171.171,39.202) (196.762,37.1996) (169.742,34.7821) (173.23,36.1422) (120,35.9981) (132,36.984) (163.22,36.9547) (117.81,37.3988) (119.829,35.0059) (200,39.1604) (182.407,34.9139) (123.946,34.0467) (207.468,39.4128) (185,38.8474) (141.372,36.326) (188.523,37.2457) (161.069,34.6878) (179.102,38.9371) (200.557,38.6557) (237.978,37.6219) (191.949,35.7317) (183.184,37.6784) (200,38.9751) (201.6,35.8402) (224,38.8618) (189.65,36.3025) (206.7,38.8175) (220.592,37.1211) (261.301,37.0613) (168.266,36.4792) (325.316,35.2105) (230.278,30.2221) (212.299,38.1885) (250.571,34.1031) (201.6,33.9597) (160.771,37.9737) (238.871,38.2023) (211.629,34.9217) (236.944,35.0849) (226.832,38.4778) (228.965,38.9217) (293.566,32.8247) (289.54,34.3614) (181.995,32.8549) (225.687,35.9894) (244.29,37.573) (240,39.0762) (233.656,25.4113) (228.003,36.5) (266.683,38.8592) (163.489,38.9208) (171.562,38.8778) (247.75,36.6418) (242.859,38.8202) (218,39.031) (243.429,38.8722) (252.434,38.7615) (329.757,36.8681) (141.771,38.3596) (270.384,29.268) (203.521,38.4652) (452.79,37.6893) (257.029,30.1745) (180.503,27.067) (266.202,28.9616) (314,34.5972) (240.72,38.9546) (282.863,32.6811) (317.922,31.5115) (265.825,31.6144) (275.272,31.6963) (231.011,37.1364) (285,32.0578) (215.711,31.1288) (231.536,39.1082) (280.646,32.1399) (265.388,28.5607) (225.762,31.9822) (365.431,28.3699) (313.143,33.695) (277.427,28.2221) (251.153,30.9201) (364,37.2705) (275.093,31.0417) (437.228,38.3559) (277.245,34.358) (247.627,29.2289) (316.844,34.8102) (218.153,34.6897) (310.635,34.0011) (303.977,30.6526) (340.515,32.1386) (315,34.7927) (257.92,35.4402) (360,37.1812) (328.251,35.4095) (316.306,35.5288) (248.896,34.3942) (332.719,32.4608) (359.1,33.8971) (280.985,34.9633) (378.829,35.1316) (318.78,36.2347) (317.931,35.6343) (335.105,36.748) (353.225,36.3131) (351,36.6479) (399.022,37.6668) (457.662,38.1229) (352.602,28.5147) (384.355,36.5294) (350,36.7418) (364.466,35.9769) (264,29.9114) (222,38.9719) (291.544,36.9462) (337.526,36.7181) (316.35,32.5235) (412.203,32.7643) (264.961,36.0869) (366.369,36.9114) (461.754,35.0621) (363.898,36.106) (472.846,37.7944) (315.414,33.8918) (389.187,26.5653) (393.673,36.6187) (363.158,37.034) (428.766,37.0408) (379.89,38.2528) (351.54,37.1814) (486,38.5176) (422.8,37.9022) (394.98,25.8562) (377.193,31.4457) (374.541,36.0673) (390,37.7068) (388.513,33.2038) (392.679,37.7873) (348.957,36.6702) (391.372,37.43) (340.29,33.1099) (410.571,37.6565) (355.925,33.928) (552.364,37.1215) (400.039,37.9946) (382.063,34.6439) (427.559,37.5543) (294.278,37.9237) (275.359,33.9914) (410.844,36.7652) (298.62,35.1061) (392.07,36.8032) (413.862,38.0077) (326.289,37.3958) (412.297,38.7717) (507.375,37.9166) (329.587,37.4226) (269.559,37.0678) (328.889,37.7807) (420,38.2274) (435.778,37.3739) (363.942,36.1751) (384.457,38.5495) (370.696,34.307) (437.555,30.32) (420.023,34.1218) (486.879,36.8401) (469.447,34.64) (335.541,29.6014) (378,37.2214) (450.562,33.6956) (255.255,36.3056) (432,38.4581) (369.808,38.8851) (353.532,38.6991) (535.502,33.3027) (489.467,31.4732) (291.519,37.8582) (372.519,34.9193) (357.335,36.5024) (531.001,29.9043) (469.243,37.369) (562.833,37.5055) (341.574,34.2525) (471.319,36.7736) (514.959,39.2492) (318.318,38.1235) (530.948,37.3092) (406.178,38.7272) (453.988,38.26) (550.188,32.8189) (489.262,38.1431) (359.507,35.7144) (400,37.7355) (490,38.6171) (475.321,26.1851) (507,38.6525) (520.024,36.6409) (306.366,36.3921) (354.91,36.2534) (524.574,33.5418) (508.151,37.7809) (392.467,37.7668)};
\addplot [mark=star,draw=none,color=violet] coordinates {(3.283500,0.166459) (3.000000,0.0918728) (5.026113,0.360122) (11.205000,0.904346) (10.482320,0.895057) (7.965500,0.696239) (13.912500,1.2008) (14.857500,1.17997) (12.264000,0.891439) (17.802000,1.64254) (20.010000,1.74202) (16.329403,1.76454) (27.135000,2.27996) (22.822242,2.11981) (26.914453,2.19116) (30.309930,2.04617) (24.839502,2.63894) (33.189109,3.19134) (23.296875,2.1946) (29.256550,2.51723) (25.927670,2.7215) (39.440000,3.25574) (28.967498,2.59254) (36.860250,3.37974) (47.528208,4.43362) (37.055559,3.80575) (44.988750,3.87385) (37.974105,3.20251) (55.805699,4.26804) (50.278125,3.38985) (64.000000,1.89576) (50.509565,4.44798) (59.962500,5.06547) (79.888032,5.92939) (62.137500,1.78716) (53.820000,4.91784) (68.330438,5.59787) (77.688000,2.25298) (61.763625,5.19012) (66.844811,3.98399) (45.168559,4.78317) (35.457750,3.01105) (61.195312,5.60542) (79.232091,5.83722) (78.390000,5.91666) (105.366000,7.38371) (81.200000,6.42722) (65.264454,5.36342) (89.072297,5.07941) (81.864210,5.87878) (88.357500,5.41914) (96.693750,2.63872) (89.437500,5.21711) (77.412510,6.83383) (116.410119,6.67778) (73.345500,5.32696) (61.873875,5.40678) (69.300000,2.93911) (100.000000,2.98556) (62.429063,5.27367) (96.642000,7.86427) (94.116968,6.13324) (139.780000,8.96645) (103.764375,8.389) (106.920000,8.91505) (89.554766,7.41813) (107.905766,8.45799) (112.664062,8.81417) (113.324750,8.97638) (98.541056,6.66223) (107.390258,3.23513) (118.965000,3.43625) (132.000000,2.92808) (112.432716,9.96444) (122.487047,9.50268) (139.104000,6.90744) (111.529756,9.51719) (127.360800,8.94086) (157.244920,10.4275) (120.000000,2.72299) (101.812482,6.70951) (88.219375,7.4223) (141.977500,3.97786) (152.558437,9.46308) (143.768625,6.61444) (168.000000,4.64068) (110.892919,7.83335) (166.897500,10.8386) (148.803480,10.63) (104.247000,3.06586) (132.961500,5.07496) (113.636355,8.98057) (155.000000,3.38396) (150.000000,4.27439) (167.034000,11.8019) (143.872116,10.7065) (171.171000,11.5927) (196.762500,12.1911) (169.742430,11.3367) (173.230000,12.5158) (120.000000,2.67508) (132.000000,3.77369) (163.220313,4.74777) (117.809737,12.4699) (119.829375,8.32813) (200.000000,5.47888) (182.407500,11.4962) (123.946372,11.0977) (207.468000,13.1889) (185.000000,4.08257) (141.372467,12.5992) (188.522539,12.8219) (161.069411,11.982) (179.101692,5.64199) (200.557500,13.8452) (237.978348,14.7084) (191.949210,12.5181) (183.183750,12.7101) (200.000000,5.51058) (201.600394,12.1216) (224.000000,4.76687) (189.649687,14.1193) (206.700000,5.52698) (220.592128,13.6989) (261.301202,14.3628) (168.265756,6.75627) (325.316250,15.5141) (230.277551,12.9343) (212.299062,13.7473) (250.570513,11.233) (201.600000,14.5085) (160.770741,9.11414) (238.871250,11.0528) (211.629000,13.7864) (236.943510,14.1868) (226.831699,14.065) (228.965000,6.32562) (293.566250,15.4853) (289.539675,7.78899) (181.994575,10.4457) (225.686957,15.0534) (244.290000,6.4603) (240.000000,6.50326) (233.656500,14.9547) (228.003072,12.6675) (266.682780,12.8934) (163.489428,11.7313) (171.562500,6.49566) (247.750000,13.2798) (242.859375,6.84846) (218.000000,4.61639) (243.429375,5.21591) (252.434063,6.86494) (329.757426,17.6065) (141.771000,11.4058) (270.384000,16.0494) (203.521500,10.0217) (452.790000,15.4713) (257.028750,12.4094) (180.502678,14.2325) (266.201912,15.4361) (314.000000,7.92761) (240.720000,14.7937) (282.862500,14.7248) (317.922412,14.9395) (265.825000,15.6396) (275.271956,16.4065) (231.011016,13.3551) (285.000000,6.99949) (215.711121,13.6757) (231.536000,14.4809) (280.645471,15.8173) (265.388062,14.8271) (225.762390,7.03497) (365.431250,16.2701) (313.142500,14.3123) (277.426758,11.2528) (251.153312,15.6428) (364.000000,6.47829) (275.093044,16.4611) (437.227875,16.6939) (277.245000,7.50927) (247.626844,14.2145) (316.844412,16.5411) (218.152861,12.6348) (310.635418,17.5273) (303.976725,16.3914) (340.515000,7.27732) (315.000000,7.59832) (257.920000,16.9142) (360.000000,7.08644) (328.251000,16.9519) (316.306250,11.3747) (248.895905,15.9761) (332.718750,17.3329) (359.100000,16.1924) (280.985435,17.8237) (378.828728,11.4577) (318.780000,15.544) (317.930589,17.0105) (335.104688,15.7752) (353.225348,8.71202) (351.000000,8.55669) (399.022059,16.2186) (457.661680,19.3384) (352.602150,16.6001) (384.355125,16.7805) (350.000000,6.9898) (364.465828,18.2549) (264.000000,5.36708) (222.000000,5.99055) (291.543750,18.0727) (337.526456,17.6715) (316.350000,17.001) (412.202810,20.8058) (264.960675,8.19888) (366.368750,16.7718) (461.754475,19.3264) (363.898062,18.3696) (472.846500,15.501) (315.414000,12.7201) (389.186674,19.1078) (393.673500,17.9413) (363.158250,15.0885) (428.766000,20.3632) (379.890000,18.5233) (351.540000,18.9969) (486.000000,8.94271) (422.800000,18.41) (394.979922,15.7325) (377.192564,18.0993) (374.540625,11.7529) (390.000000,9.56375) (388.513125,18.1776) (392.679375,17.8423) (348.957000,19.5463) (391.371750,17.8128) (340.290000,17.1232) (410.571000,17.9931) (355.925281,16.8588) (552.364154,15.4687) (400.038750,9.30865) (382.063500,19.67) (427.559147,20.2896) (294.277500,9.65379) (275.358677,16.6721) (410.844068,19.7232) (298.620197,15.3767) (392.070000,20.1277) (413.861871,18.3699) (326.288813,12.9192) (412.297169,21.0873) (507.375000,17.7744) (329.587002,19.4176) (269.559404,19.3132) (328.889078,15.9196) (420.000000,7.12231) (435.778125,16.3195) (363.941865,18.0706) (384.456806,19.4778) (370.696500,18.9629) (437.555436,20.3554) (420.022969,16.9812) (486.878438,18.3548) (469.447250,19.2358) (335.541497,16.8774) (378.000000,6.86684) (450.562500,20.0526) (255.255000,8.29336) (432.000000,10.1267) (369.808000,18.354) (353.531656,20.1166) (535.501824,18.7762) (489.466980,19.7595) (291.519250,17.4291) (372.519000,13.9549) (357.335000,17.8216) (531.001170,21.3407) (469.243125,20.2938) (562.832559,20.5084) (341.573863,12.516) (471.318938,19.9378) (514.959375,20.0135) (318.318000,20.4435) (530.948014,22.654) (406.177813,15.5715) (453.988104,20.2841) (550.188000,22.3835) (489.261596,20.6018) (359.507369,19.6106) (400.000000,7.58555) (490.000000,8.87705) (475.321031,21.7713) (507.000000,10.8455) (520.023750,22.8191) (306.365976,12.4987) (354.909586,21.3989) (524.574192,21.6757) (508.151437,21.415) (392.467383,17.182)};
\legend{BLIS, BSMTC, TTT}
\end{axis}
\end{tikzpicture}

}

\caption{\label{fig:rand-rankk}Performance of matrix multiplication and randomly
generated tensor contractions for rank-$k$ update matrix/tensor shapes
on a Xeon E5-2690 v3 processor.}
\end{figure}

In order to assess the overall performance of BSMTC compared to both
matrix multiplication (of similar size and shape) and to TTT, we measured
the performance of randomly generated tensor contractions and corresponding
matrix multiplications for a range of overall tensor/matrix sizes
of two shapes: square ($m=n=k$) and rank-$k$ update ($m=n\gg k$).
These problem sizes and shapes span a reasonable range of possible
computations along the orthogonal axes of total problem size and communication
vs. computation bound problems, both factors that are expected to
affect the relative performance of BSMTC compared to TTT. The square
problem sizes investigated range from 20 to 2000 for single-core runs
and from 50 to 5000 for multi-core runs (giving a problem $\sim16$
times as large on twelve cores), while the $k$ length for the rank-$k$
update cases ranges from 5-500 in all cases, with $m=n=4000$ on one
core and $m=n=16000$ on twelve cores (again a $16\times$ increase
in problem size). For each of the problem sizes expressed as a matrix
multiplication (i.e. in terms of $m$, $n$, and $k$), we randomly
generate three similarly-shaped tensor contractions. For each matrix
dimension, we randomly choose from one to three tensor dimensions
for the corresponding bundle, where the product of the tensor dimensions
is close to the original matrix dimension. The order of the dimensions
in each tensor is then randomly permuted. In order to plot the tensor
contraction results in a fashion consistent with the prescribed matrix
multiplication sizes, ``effective'' matrix lengths for the generated
tensor contractions are determined from the actual number $f$ of
FLOPs performed. For square problem sizes we set $\tilde{m}=\tilde{n}=\tilde{k}=(\frac{f}{2})^{1/3}$,
and for the rank-$k$ update cases we set $\tilde{k}=\frac{f}{2mn}$
where $m$ and $n$ are the fixed matrix dimensions. The performance
results for matrix multiplication (using BLIS), and for tensor contraction
with BSMTC and TTT are given in \figref{rand-square} and \figref{rand-rankk}.

The performance of the BSMTC algorithm is very close to that of raw
matrix multiplication for the majority of tensor shapes. For the parallel
rank-$k$ update problems, some tensor shapes lead to reduced performance
with BSMTC. These shapes belong to one of two classes, (1) tensors
with a very small leading edge length, which inhibits the performance
benefit of the blocked scatter vector (i.e. all operations must use
the full scatter vector), and (2) contractions where stride-1 access
cannot be obtained in all three tensors simultaneously. Operations
in both of these classes are marginally affected in the computation
bound regime, but are disproportionately penalized when communication
(memory accesses) are the limiting factor. It may be possible to combine
BSMTC with the technique of dimension sub-division used in \cite{gett}
to improve locality in the packing kernel (this would essentially
yield a 3-D packing kernel) to address class (2).

The TTT results meet the performance of matrix multiplication and
BSMTC only for square tensor shapes on a single core. Moving either
to multi-core or to more communication bound tensor shapes (such as
in rank-$k$ update) results in a significant slow-down. For parallel
square tensor contractions, TTT only achieves $\sim\frac{1}{4}$ to
$\frac{3}{4}$ the performance of BSMTC, becoming competitive only
for very large matrix sizes. Similarly TTT is $\sim\frac{1}{2}$ as
fast as BSMTC on average for rank-$k$ updates on a single cores,
dropping to $\sim\frac{1}{4}$ on average in parallel.

In parallel, BLIS achieves approximately 90\% parallel weak-scalability
(i.e. the per-core parallel performance is 90\% of the single-core
performance), which is nearly identical to the scaling of MKL. The
scalability for BSMTC is only slightly less, possibly due to load
imbalance stemming from edge cases which must use the full scatter
vector. TTT, as might be expected, shows significantly lower scalability,
since the tensor transposition step is inherently bandwidth limited.
The performance variability is also reduced reduced in parallel (except
for an evident bifurcation of the parallel results for rank-$k$,
possibly an artifact of the parallelization scheme employed by MATLAB),
since a large number of threads may reach the bandwidth limit fairly
easily while a single core requires a highly efficient and tuned tensor
transpose kernel to do the same. Thus, while more efficient tensor
transposition may be beneficial to TTT on a single core, the benefit
in parallel may be somewhat more limited, especially compared to the
performance gains evidenced by BSMTC.

\subsection{Explicit Tensor Contractions\label{sec:benchmark}}

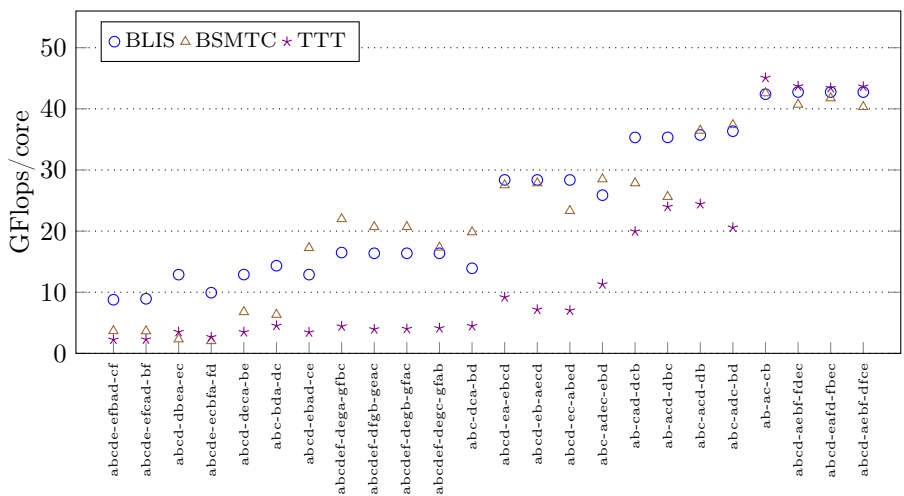
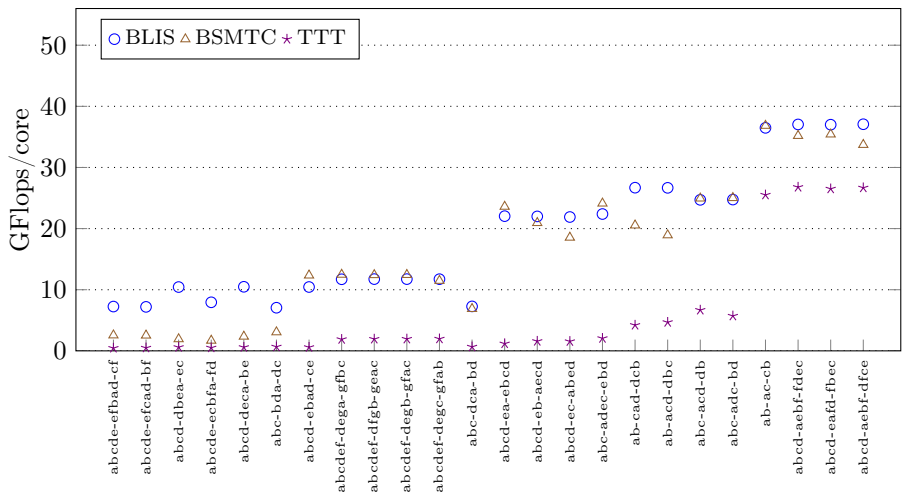
\begin{figure}
\begin{centering}
\subfloat[Single-core performance.]{\begin{centering}
\thispagestyle{empty}
\begin{tikzpicture}
\pgfplotsset {grid style={dotted,color=black}}
\begin{axis}[
x post scale=1.6,
y post scale=0.8,
ymajorgrids=true,
enlarge x limits=0.05,
ymin=0,
ymax=56,
ylabel={GFlops/core},
ylabel shift=-5,
ytick distance=10,
xticklabels={{abcde-efbad-cf}, {abcde-efcad-bf}, {abcd-dbea-ec}, {abcde-ecbfa-fd}, {abcd-deca-be}, {abc-bda-dc}, {abcd-ebad-ce}, {abcdef-dega-gfbc}, {abcdef-dfgb-geac}, {abcdef-degb-gfac}, {abcdef-degc-gfab}, {abc-dca-bd}, {abcd-ea-ebcd}, {abcd-eb-aecd}, {abcd-ec-abed}, {abc-adec-ebd}, {ab-cad-dcb}, {ab-acd-dbc}, {abc-acd-db}, {abc-adc-bd}, {ab-ac-cb}, {abcd-aebf-fdec}, {abcd-eafd-fbec}, {abcd-aebf-dfce}},
xtick=data,
only marks,
x tick label style={rotate=90,font=\tiny},
xtick pos=left,
legend style={legend pos=north west,font=\footnotesize},
legend columns=3
]
\addplot [mark=o,draw=none,color=blue] coordinates {(0,8.77823) (1,8.92601) (2,12.8797) (3,9.9295) (4,12.8895) (5,14.342) (6,12.8827) (7,16.4908) (8,16.3714) (9,16.3718) (10,16.3736) (11,13.9174) (12,28.3748) (13,28.3729) (14,28.365) (15,25.8786) (16,35.3242) (17,35.335) (18,35.7231) (19,36.3438) (20,42.4012) (21,42.7526) (22,42.748) (23,42.7402)};
\addplot [mark=triangle,draw=none,color=brown!80!black] coordinates {(0,3.67145) (1,3.63349) (2,2.31697) (3,2.02699) (4,6.77274) (5,6.33973) (6,17.2525) (7,21.9863) (8,20.6799) (9,20.6898) (10,17.3284) (11,19.8354) (12,27.5242) (13,27.8466) (14,23.3171) (15,28.5081) (16,27.8475) (17,25.6178) (18,36.4539) (19,37.3686) (20,42.568) (21,40.7005) (22,41.747) (23,40.3194)};
\addplot [mark=star,draw=none,color=violet] coordinates {(0,2.245203) (1,2.278907) (2,3.505533) (3,2.660330) (4,3.475592) (5,4.524059) (6,3.434642) (7,4.418373) (8,3.943019) (9,4.005366) (10,4.123714) (11,4.464822) (12,9.179275) (13,7.157137) (14,7.015273) (15,11.296876) (16,19.945672) (17,23.970287) (18,24.420627) (19,20.585500) (20,45.081610) (21,43.694346) (22,43.415711) (23,43.644205)};
\legend{BLIS, BSMTC, TTT}
\end{axis}
\end{tikzpicture}
\par\end{centering}
}
\par\end{centering}
\begin{centering}
\subfloat[Multi-core performance (twelve cores).]{\begin{centering}
\thispagestyle{empty}
\begin{tikzpicture}
\pgfplotsset {grid style={dotted,color=black}}
\begin{axis}[
x post scale=1.6,
y post scale=0.8,
ymajorgrids=true,
enlarge x limits=0.05,
ymin=0,
ymax=56,
ylabel={GFlops/core},
ylabel shift=-5,
only marks,
ytick distance=10,
xticklabels={{abcde-efbad-cf}, {abcde-efcad-bf}, {abcd-dbea-ec}, {abcde-ecbfa-fd}, {abcd-deca-be}, {abc-bda-dc}, {abcd-ebad-ce}, {abcdef-dega-gfbc}, {abcdef-dfgb-geac}, {abcdef-degb-gfac}, {abcdef-degc-gfab}, {abc-dca-bd}, {abcd-ea-ebcd}, {abcd-eb-aecd}, {abcd-ec-abed}, {abc-adec-ebd}, {ab-cad-dcb}, {ab-acd-dbc}, {abc-acd-db}, {abc-adc-bd}, {ab-ac-cb}, {abcd-aebf-fdec}, {abcd-eafd-fbec}, {abcd-aebf-dfce}},
xtick=data,
x tick label style={rotate=90,font=\tiny},
xtick pos=left,
legend style={legend pos=north west,font=\footnotesize},
legend columns=3
]
\addplot [mark=o,draw=none,color=blue] coordinates {(0,7.24745) (1,7.19654) (2,10.4385) (3,7.94283) (4,10.4812) (5,7.05511) (6,10.4458) (7,11.711) (8,11.7384) (9,11.7366) (10,11.7253) (11,7.27176) (12,22.0354) (13,21.9937) (14,21.8953) (15,22.3824) (16,26.6921) (17,26.6673) (18,24.7038) (19,24.7479) (20,36.4804) (21,37.0374) (22,36.9967) (23,37.0711)};
\addplot [mark=triangle,draw=none,color=brown!80!black] coordinates {(0,2.55516) (1,2.52731) (2,1.93098) (3,1.68672) (4,2.34865) (5,3.06153) (6,12.3495) (7,12.4761) (8,12.4146) (9,12.4398) (10,11.479) (11,6.87438) (12,23.6) (13,20.9347) (14,18.541) (15,24.1243) (16,20.5351) (17,18.9286) (18,24.9435) (19,25.019) (20,36.8373) (21,35.1822) (22,35.4236) (23,33.7126)};
\addplot [mark=star,draw=none,color=violet] coordinates {(0,0.453099) (1,0.487117) (2,0.592353) (3,0.505951) (4,0.589194) (5,0.666707) (6,0.584637) (7,1.88379) (8,1.92886) (9,1.93919) (10,1.96667) (11,0.655881) (12,1.17799) (13,1.57712) (14,1.54637) (15,2.052) (16,4.20438) (17,4.67433) (18,6.6704) (19,5.71434) (20,25.5069) (21,26.7765) (22,26.5055) (23,26.6782)};
\legend{BLIS, BSMTC, TTT}
\end{axis}
\end{tikzpicture}
\par\end{centering}
}
\par\end{centering}
\caption{\label{fig:bench}Performance of a variety of tensor contractions
and the equivalent matrix multiplications on a Xeon E5-2690 v3 processor.}
\end{figure}

In addition to randomly generated tensor contractions, we have also
measured the performance of BSMTC and TTT for a set of tensor contractions
from the Tensor Contraction Benchmark \cite{tensor_benchmark}. The
performance for both algorithms was measured for each tensor contraction
both on a single core and across all twelve cores of the processor
(using the same tensor size). The performance of an equivalent matrix
multiplication was also measured for each contraction as a reference.
The results for the benchmark are collected in \figref{bench}. The
specific contraction is identified by the index string, which lists
the tensor indices of each tensor in the order $\mathscr{C}-\mathscr{A}-\mathscr{B}$.
Thus, the string $abc-adec-ebd$ denotes the contraction $\mathscr{C}_{abc}\coloneqq\mathscr{A}_{adec}\cdot\mathscr{B}_{ebd}$.

The tensor contractions are arranged from communication bound on the
left to computation bound on the right. In the single-core computation
bound cases, as for randomly generated square tensor contractions,
the performance of BLIS, BSMTC, and TTT is similar and very close
to the peak performance of the machine. As the contractions become
more communication bound, the absolute performance for all algorithms
drops, with BLIS gradually dropping from $\sim40$ GFLOPs to $\sim10$
GFLOPs. BSMTC performance is broadly similar, except for the left-most
six cases. TTT, however, shows a much sharper drop-off in performance,
very quickly dropping to $\sim20$ GFLOPs and then shortly thereafter
to $<10$ GFLOPs while BSMTC still maintains close to 30 GFLOPs. While
BSMTC performance is greatly improved over that for TTT, there are
some instances where it does not recover the full performance of matrix
multiplication. On inspecting the tensor indices for these contractions,
for example $abcde-efcad-bf$, one may note that the tensor $\mathscr{A}_{efcad}$
cannot be packed with stride-1 access (since the $e$ dimension has
larger stride than $a$ in the $\mathscr{C}$ tensor, which always
takes priority). Since packing this tensor requires a significant
portion of the running time, using an inefficient access pattern is
especially detrimental. As mentioned previously, this performance
bottleneck may be avoided by using a more complicated packing kernel.
However, BSMTC also exceeds the performance of BLIS in several instances.
This is likely due to the particular transpose variant of matrix multiplication
employed (all BLIS results use the ``No transpose/No transpose''
variant), which may be less optimal than that used inside BSMTC.

The parallel results show similar trends as the single-core results,
but to a much more extreme degree. While BLIS and BSMTC performance
is only lightly impacted by strongly scaling to twelve cores, TTT
can only surpass $\sim5$ GFLOPs/core for the large, square tensor
shapes. The speedup of BSMTC over TTT in the single-core case ranges
from $\sim0.7$ ($abcd-dbea-ec$) to $\sim5$ ($abcdef-dfgb-geac$),
while in the multi-core case it ranges from $\sim1.3$ ($ab-ac-cb$
and following) to 21.1 ($abcd-ebad-ce$). The speedups for the six-dimensional
tensor cases are especially exciting as they represent critical contractions
encountered in the popular CCSD(T) quantum chemistry method \cite{raghavachari_fifth-order_1989,apra_efficient_2014}.

\section{Summary and Conclusions}

We have presented two novel mappings from a general tensor data layout
to a matrix layout, which allow for tensor elements to be accessed
in-place from within matrix-oriented computational kernels. We have
shown how kernels using these mappings within the BLIS approach to
matrix multiplication produce new tensor contractions algorithms which
we denote Scatter-Matrix Tensor Contraction (SMTC) and Block-Scatter-Matrix
Tensor Contraction (BSMTC). These approaches achieve an efficiency
uniformly higher than that of the traditional TTGT approach as implemented
in the MATLAB Tensor Toolbox, approaching that of matrix multiplication
using the BLIS framework. Our implementations of these algorithms
also achieves excellent parallel scalability when using multithreading.

The BSMTC algorithm exhibits performance very close to that of matrix
multiplication across a wide variety of tensor shapes and sizes, in
sequential as well as in multithreaded execution, while also avoiding
the workspace requirements of TTGT, we conclude that these algorithms
should be considered as high-performance alternatives to existing
tensor contraction algorithms. The efficiency and simplicity of these
algorithms also highlights the utility and flexibility of the BLIS
approach to matrix multiplication.

\section*{Source Code Availability}

The BSMTC tensor contraction algorithm has been implemented in the
TBLIS (Tensor-Based Library Instantiation Software) framework, which
is available under a BSD license at https://\foreignlanguage{australian}{github}.com/devinamatthews/tblis.

\section*{Acknowledgments}

DAM would like to thank Field Van Zee and Tyler Smith for many helpful
discussion about BLIS, matrix multiplication, multithreading, and
many other topics, and Prof. Robert van de Geijn for a critical reading
of the manuscript and many useful suggestions. He would also like
to thank the Arnold and Mabel Beckman Foundation for support as an
Arnold O. Beckman Postdoctoral Fellow, and gratefully acknowledges
funding from the National Science Foundation under grant number ACI-1148125/1340293
and from Intel Corp. through an Intel Parallel Computing Center grant.
The authors acknowledge the Texas Advanced Computing Center (TACC)
at The University of Texas at Austin for providing HPC resources that
have contributed to the research results reported within this paper.
URL: http://www.tacc.utexas.edu 

\bibliographystyle{plain}
\bibliography{tensor}

\begin{thebibliography}{10}

\bibitem{apra_efficient_2014}
E.~Apr{\`a} , M.~Klemm, and K.~Kowalski.
\newblock Efficient {Implementation} of {Many}-body {Quantum} {Chemical}
  {Methods} on the {Intel\textregistered} {Xeon} {Phi\texttrademark}
  {Coprocessor}.
\newblock In {\em Proceedings of the {International} {Conference} for {High}
  {Performance} {Computing}, {Networking}, {Storage} and {Analysis}}, {SC} '14,
  pages 674--684, Piscataway, NJ, USA, 2014. IEEE Press.

\bibitem{ttoolbox}
B.~W. Bader and T.~G. Kolda.
\newblock Algorithm 862: {MATLAB} {tensor} {classes} for {fast} {algorithm}
  {prototyping}.
\newblock {\em ACM Trans. Math. Softw.}, 32(4):635--653, 2006.

\bibitem{bartlett_coupled-cluster_2007}
R.~J. Bartlett and M.~Musia{\l}.
\newblock Coupled-cluster theory in quantum chemistry.
\newblock {\em Rev. Mod. Phys.}, 79(1):291--352, 2007.

\bibitem{bto}
G.~Belter, E.~R. Jessup, I.~Karlin, and J.~G. Siek.
\newblock Automating the {generation} of {composed} {linear} {algebra}
  {kernels}.
\newblock In {\em Proceedings of the {Conference} on {High} {Performance}
  {Computing} {Networking}, {Storage} and {Analysis}}, {SC} '09, pages
  59:1--59:12, New York, NY, USA, 2009. ACM.

\bibitem{tiledarray1}
J.~A. Calvin, C.~A. Lewis, and E.~F. Valeev.
\newblock Scalable {task}-based {algorithm} for {multiplication} of
  {block}-rank-sparse {matrices}.
\newblock In {\em Proceedings of the 5th {Workshop} on {Irregular}
  {Applications}: {Architectures} and {Algorithms}}, {IA}$^{\textrm{3}}$ '15,
  pages 4:1--4:8, New York, NY, USA, 2015. ACM.

\bibitem{tiledarray2}
J.~A. Calvin and E.~F. Valeev.
\newblock Task-{based} {algorithm} for {matrix} {multiplication}: {A} {step}
  {towards} {block}-{sparse} {tensor} {computing}.
\newblock {\em arXiv:1504.05046 [cs]}, 2015.

\bibitem{slice1}
E.~Di~Napoli, D.~Fabregat-Traver, G.~Quintana-Orti, and P.~Bientinesi.
\newblock Towards an efficient use of the {BLAS} library for multilinear tensor
  contractions.
\newblock {\em Appl. Math. Comput.}, 235:454--468, 2014.

\bibitem{blas3}
J.~J. Dongarra, J.~Du~Croz, S.~Hammarling, and I.~S. Duff.
\newblock A {set} of {level} 3 {basic} {linear} {algebra} {subprograms}.
\newblock {\em ACM Trans. Math. Softw.}, 16(1):1--17, 1990.

\bibitem{blas2}
J.~J. Dongarra, J.~Du~Croz, S.~Hammarling, and R.~J. Hanson.
\newblock An {extended} {set} of {FORTRAN} {basic} {linear} {algebra}
  {subprograms}.
\newblock {\em ACM Trans. Math. Softw.}, 14(1):1--17, 1988.

\bibitem{libtensor}
E.~Epifanovsky, M.~Wormit, T.~Kus, A.~Landau, D.~Zuev, K.~Khistyaev,
  P.~Manohar, I.~Kaliman, A.~Dreuw, and A.~I. Krylov.
\newblock New implementation of high-level correlated methods using a general
  block-tensor library for high-performance electronic structure calculations.
\newblock {\em J. Comp. Chem.}, 34(26):2293--2309, 2013.

\bibitem{goto1}
K.~Goto and R.~A. van~de Geijn.
\newblock Anatomy of {high}-performance {matrix} {multiplication}.
\newblock {\em ACM Trans. Math. Softw.}, 34(3):12:1--12:25, 2008.

\bibitem{goto2}
K.~Goto and R.~A. van~de Geijn.
\newblock High-performance {implementation} of the {level}-3 {BLAS}.
\newblock {\em ACM Trans. Math. Softw.}, 35(1):4:1--4:14, 2008.

\bibitem{Eigen}
G.~Guennebaud, B.~Jacob, et~al.
\newblock Eigen v3, 2010.
\newblock http://eigen.tuxfamily.org.

\bibitem{hanrath_efficient_2010}
M.~Hanrath and A.~Engels-Putzka.
\newblock An efficient matrix-matrix multiplication based antisymmetric tensor
  contraction engine for general order coupled cluster.
\newblock {\em J. Chem. Phys.}, 133(6):064108, 2010.

\bibitem{hartono_performance_2009}
A.~Hartono, Q.~Lu, T.~Henretty, S.~Krishnamoorthy, H.~Zhang, G.~Baumgartner,
  D.~E. Bernholdt, M.~Nooijen, R.~Pitzer, J.~Ramanujam, and P.~Sadayappan.
\newblock Performance {optimization} of {tensor} {contraction} {expressions}
  for {many}-{body} {methods} in {quantum} {chemistry}.
\newblock {\em J. Phys. Chem. A}, 113(45):12715--12723, 2009.

\bibitem{tce}
S.~Hirata.
\newblock Tensor {contraction} {engine}: {Abstraction} and {automated}
  {parallel} {implementation} of {configuration}-{interaction},
  {coupled}-{cluster}, and {many}-{body} {perturbation} {theories}.
\newblock {\em J. Phys. Chem. A}, 107(46):9887--9897, 2003.

\bibitem{FLAWN79}
J.~Huang, T.~M. Smith, G.~M. Henry, and R.~A. van~de Geijn.
\newblock {Strassen's} algorithm reloaded.
\newblock In {\em Proceedings of the {Conference} on {High} {Performance}
  {Computing} {Networking}, {Storage} and {Analysis}}, {SC} '16, pages
  59:1--59:12, New York, NY, USA, 2016. ACM.

\bibitem{kolda_tensor_2009}
T.~Kolda and B.~Bader.
\newblock Tensor {decompositions} and {applications}.
\newblock {\em SIAM Rev.}, 51(3):455--500, 2009.

\bibitem{kroonenberg_applied_2008}
P.~M. Kroonenberg.
\newblock {\em Applied {multiway} {data} {analysis}}.
\newblock John Wiley \& Sons, 2008.

\bibitem{blas1}
C.~L. Lawson, R.~J. Hanson, D.~R. Kincaid, and F.~T. Krogh.
\newblock Basic {linear} {algebra} {subprograms} for {FORTRAN} {usage}.
\newblock {\em ACM Trans. Math. Softw.}, 5(3):308--323, 1979.

\bibitem{slice3}
J.~Li, C.~Battaglino, I.~Perros, J.~Sun, and R.~Vuduc.
\newblock An {input}-adaptive and {in}-place {approach} to {dense}
  {tensor}-times-matrix {multiply}.
\newblock In {\em Proceedings of the {International} {Conference} for {High}
  {Performance} {Computing}, {Networking}, {Storage} and {Analysis}}, {SC} '15,
  pages 76:1--76:12, New York, NY, USA, 2015. ACM.

\bibitem{lyakh_efficient_2015}
Dmitry~I. Lyakh.
\newblock An efficient tensor transpose algorithm for multicore {CPU}, {Intel}
  {Xeon} {Phi}, and {NVidia} {Tesla} {GPU}.
\newblock {\em Comput. Phys. Commun.}, 189:84--91, 2015.

\bibitem{slice_gpu1}
W.~Ma, S.~Krishnamoorthy, O.~Villa, K.~Kowalski, and G.~Agrawal.
\newblock Optimizing tensor contraction expressions for hybrid {CPU}-{GPU}
  execution.
\newblock {\em Cluster Comput.}, 16(1):131--155, 2011.

\bibitem{dxter}
B.~Marker, D.~Batory, and R.~A. van~de Geijn.
\newblock A case study in mechanically deriving dense linear algebra code.
\newblock {\em Int. J. High Perform. C.}, 27(4):440--453, 2013.

\bibitem{matthews_non-orthogonal_2015}
D.~A. Matthews and J.~F. Stanton.
\newblock Non-orthogonal spin-adaptation of coupled cluster methods: {A} new
  implementation of methods including quadruple excitations.
\newblock {\em J. Chem. Phys.}, 142(6):064108, 2015.

\bibitem{slice2}
E.~Peise, D.~Fabregat-Traver, and P.~Bientinesi.
\newblock On the {performance} {prediction} of {BLAS}-based {tensor}
  {contractions}.
\newblock In S.~A. Jarvis, S.~A. Wright, and S.~D. Hammond, editors, {\em High
  {Performance} {Computing} {Systems}. {Performance} {Modeling},
  {Benchmarking}, and {Simulation}}, number 8966 in Lecture {Notes} in
  {Computer} {Science}, pages 193--212. Springer International Publishing,
  2014.
\newblock DOI: 10.1007/978-3-319-17248-4\_10.

\bibitem{raghavachari_fifth-order_1989}
K.~Raghavachari, G.~W. Trucks, J.~A. Pople, and M.~Head-Gordon.
\newblock A fifth-order perturbation comparison of electron correlation
  theories.
\newblock {\em Chem. Phys. Lett.}, 157:479--483, 1989.

\bibitem{smilde_multi-way_2005}
A.~Smilde, R.~Bro, and P.~Geladi.
\newblock {\em Multi-way {analysis}: {Applications} in the {chemical}
  {sciences}}.
\newblock John Wiley \& Sons, 2005.

\bibitem{blis-multi}
T.~M. Smith, R.~A. van~de Geijn, M.~Smelyanskiy, J.~R. Hammond, and F.~G.
  Van~Zee.
\newblock Anatomy of {high}-{performance} {many}-{threaded} {matrix}
  {multiplication}.
\newblock In {\em Parallel and {Distributed} {Processing} {Symposium}, 2014
  {IEEE} 28th {International}}, pages 1049--1059, 2014.

\bibitem{ctf}
E.~Solomonik, D.~Matthews, J.~R. Hammond, J.~F. Stanton, and J.~Demmel.
\newblock A massively parallel tensor contraction framework for coupled-cluster
  computations.
\newblock {\em J. Par. Dist. Comp.}, 74(12):3176--3190, 2014.

\bibitem{tensor_benchmark}
P.~Springer and P.~Bientinesi.
\newblock {Tensor} {Contraction} {Benchmark} {v0.1},
  https://github.com/hpac/tccg/tree/master/benchmark.

\bibitem{gett}
P.~Springer and P.~Bientinesi.
\newblock Design of a high-performance {GEMM}-like tensor-tensor
  multiplication.
\newblock {\em arXiv:1607.00145 [cs]}, 2016.

\bibitem{ttc}
P.~Springer, J.~R. Hammond, and P.~Bientinesi.
\newblock {TTC}: {A} high-performance {compiler} for {tensor} {transpositions}.
\newblock {\em arXiv:1603.02297 [cs]}, 2016.
\newblock submitted to ACM TOMS.

\bibitem{DPD}
J.~F. Stanton, J.~Gauss, J.~D. Watts, and R.~J. Bartlett.
\newblock A direct product decomposition approach for symmetry exploitation in
  many-body methods. {I}. {Energy} calculations.
\newblock {\em J. Chem. Phys.}, 94(6):4334--4345, 1991.

\bibitem{NumPy}
S.~van~de Walt, S.~C. Colbert, and G.~Varoquaux.
\newblock The {NumPy} {array}: {A} {structure} for {efficient} {numerical}
  {computation}.
\newblock {\em Comput. Sci. Eng.}, 13(2):22--30, 2011.

\bibitem{blis2}
F.~G. Van~Zee, T.~M. Smith, B.~Marker, T.~M. Low, R.~A. van~de Geijn, F.~D.
  Igual, M.~Smelyanskiy, X.~Zhang, V.~A. Kistler, J.~A. Gunnels, and
  L.~Killough.
\newblock The {BLIS} {framework}: {Experiments} in {portability}.
\newblock {\em ACM Trans. Math. Softw.}
\newblock in press.

\bibitem{blis1}
F.~G. Van~Zee and R.~A. van~de Geijn.
\newblock {BLIS}: {A} {framework} for {rapidly} {instantiating} {BLAS}
  {functionality}.
\newblock {\em ACM Trans. Math. Softw.}, 41(3):14:1--14:33, 2015.

\bibitem{vasilescu_multilinear_2002}
M.~A.~O. Vasilescu and D.~Terzopoulos.
\newblock Multilinear {analysis} of {image} {ensembles}: {TensorFaces}.
\newblock In A.~Heyden, G.~Sparr, M.~Nielsen, and P.~Johansen, editors, {\em
  Computer {Vision} {ECCV} 2002}, number 2350 in Lecture {Notes} in {Computer}
  {Science}, pages 447--460. Springer Berlin Heidelberg, 2002.
\newblock DOI: 10.1007/3-540-47969-4\_30.

\bibitem{Blitz}
T.~L. Veldhuizen.
\newblock Arrays in {Blitz}++.
\newblock In D.~Caromel, R.~R. Oldehoeft, and M.~Tholburn, editors, {\em
  Computing in {Object}-{Oriented} {Parallel} {Environments}}, number 1505 in
  Lecture {Notes} in {Computer} {Science}, pages 223--230. Springer Berlin
  Heidelberg, 1998.
\newblock DOI: 10.1007/3-540-49372-7\_24.

\bibitem{openblas2}
Q.~Wang, X.~Zhang, Y.~Zhang, and Q.~Yi.
\newblock {AUGEM}: {Automatically} {generate} {high} {performance} {dense}
  {linear} {algebra} {kernels} on x86 {CPUs}.
\newblock In {\em Proceedings of the {International} {Conference} on {High}
  {Performance} {Computing}, {Networking}, {Storage} and {Analysis}}, {SC} '13,
  pages 25:1--25:12, New York, NY, USA, 2013. ACM.

\bibitem{chenhan}
C.~D. Yu, J.~Huang, W.~Austin, B.~Xiao, and G.~Biros.
\newblock Performance {optimization} for the {K}-nearest {neighbors} {kernel}
  on x86 {architectures}.
\newblock In {\em Proceedings of the {International} {Conference} for {High}
  {Performance} {Computing}, {Networking}, {Storage} and {Analysis}}, {SC} '15,
  pages 7:1--7:12, New York, NY, USA, 2015. ACM.

\bibitem{openblas1}
X.~Zhang, Q.~Wang, and Y.~Zhang.
\newblock Model-driven {level} 3 {BLAS} {performance} {optimization} on
  {Loongson} 3a {processor}.
\newblock In {\em 2012 {IEEE} 18th {International} {Conference} on {Parallel}
  and {Distributed} {Systems} ({ICPADS})}, pages 684--691, 2012.

\end{thebibliography}

\end{document}